\documentclass[preprint]{aastex6}
\usepackage{graphicx}
\usepackage{natbib}
\usepackage{array}
\usepackage{xcolor}
\usepackage[caption=false]{subfig}
\usepackage{natbib}
\usepackage{amssymb}

\begin{document}
\title{H$_2$ Fluorescence in M dwarf Systems: A Stellar Origin}

\author{Nicholas Kruczek\altaffilmark{1}}
\author{Kevin France\altaffilmark{1}}
\author{William Evonosky\altaffilmark{2}}
\author{R. O. Parke Loyd\altaffilmark{1}}
\author{Allison Youngblood\altaffilmark{1}}
\author{Aki Roberge\altaffilmark{3}}
\author{Robert A. Wittenmyer\altaffilmark{4,5}}
\author{John T. Stocke\altaffilmark{6}}
\author{Brian Fleming\altaffilmark{1}}
\author{Keri Hoadley\altaffilmark{1}}

\affil{$^{\rm 1}$ Laboratory for Atmospheric and Space Physics, University of Colorado, 600 UCB, Boulder, CO 80309, USA; nicholas.kruczek@colorado.edu \\ 
$^{\rm 2}$ Department of Physics, University of South Florida, Tampa, FL 33620, USA \\
$^{\rm 3}$ Exoplanets and Stellar Astrophysics Laboratory, NASA Goddard Space Flight Center, Greenbelt, MD 20771, USA \\
$^{\rm 4}$ University of Southern Queensland, Computational Engineering and Science Research Centre, Toowoomba, Queensland 4350, Australia\\
$^{\rm 5}$ School of Physics and Australian Centre for Astrobiology, University of New South Wales, Sydney 2052, Australia \\
$^{\rm 6}$ Center for Astrophysics and Space Astronomy, University of Colorado, 389 UCB, Boulder, CO 80309, USA
}

\begin{abstract}
Observations of molecular hydrogen (H$_2$) fluorescence are a potentially useful tool for measuring the H$_2$ abundance in exoplanet atmospheres. This emission was previously observed in M$\;$dwarfs with planetary systems. However, low signal-to-noise prevented a conclusive determination of its origin. Possible sources include exoplanetary atmospheres, circumstellar gas disks, and the stellar surface. We use observations from the ``Measurements of the Ultraviolet Spectral Characteristics of Low-mass Exoplanet Host Stars" (MUSCLES) Treasury Survey to study H$_2$ fluorescence in M$\;$dwarfs. We detect fluorescence in {\it Hubble Space Telescope} spectra of 8/9 planet-hosting and 5/6 non-planet-hosting M$\;$dwarfs. The detection statistics, velocity centroids, and line widths of the emission suggest a stellar origin. We calculate H$_2$-to-stellar-ion flux ratios to compare flux levels between stars. For stars with planets, we find an average ratio of 1.7$\,\pm\,$0.9 using the fluxes of the brightest H$_2$ feature and two stellar \ion{C}{4} lines. This is compared to 0.9$\,\pm\,$0.4 for stars without planets, showing that the planet-hosting M$\;$dwarfs do not have significant excess H$_{2}$ emission. This claim is supported by the direct FUV imaging of GJ 832, where no fluorescence is observed at the expected star-planet separation. Additionally, the 3-$\sigma$ upper limit of 4.9$\,\times\,$10$^{-17}$ erg$\;$cm$^{-2}\;$s$^{-1}$ from these observations is two orders of magnitude below the spectroscopically-observed H$_2$ flux. We constrain the location of the fluorescing H$_2$ using 1D radiative transfer models and find that it could reside in starspots or a $\sim$2500-3000$\;$K region in the lower chromosphere. The presence of this emission could complicate efforts to quantify the atmospheric abundance of H$_2$ in exoplanets orbiting M$\;$dwarfs.
\end{abstract}

\section{INTRODUCTION}
Exoplanet transits enable the study of planetary atmospheric composition via transmission spectroscopy. Future space-based missions, like the {\it James Webb Space Telescope}, will be capable of characterizing transiting exoplanet atmospheres, with a few dozen short period planets likely to be thoroughly studied (\citealt{Cowan15}). To properly interpret the results, it will be necessary to account for the stellar environments around these planets, since the host stars can influence the observed atmospheric features. For example, the smaller stellar radii (R$_{*}$) of M dwarfs make them favorable targets for transmission spectroscopy because the ratio of the radius of a transiting planet to R$_{*}$ is larger, meaning that planetary transits are more pronounced and therefore atmospheric absorption is easier to detect. However, the high energy irradiance and temporal variability from these stars have only recently been refined by observation (\citealt{Hawley96,West04,Welsh07,Walkowicz11,France13,Loyd14}). An improved understanding of the ultraviolet (UV) spectrum of M dwarfs is important for understanding the variations in the transit depths of UV-sensitive atoms and molecules within the planetary atmosphere.

The UV bandpass (100-3200~\AA) contains the strongest molecular transitions of abundant species, such as H$_2$, CO, H$_2$O, and O$_2$, that are important to planetary atmospheric chemistry. H$_2$ acts as a sink for positive ions, changing the predicted electron density (\citealt{Cravens87}), and as a shield, protecting other molecules from dissociation (\citealt{Kim94}). Therefore, an accurate estimate of the H$_2$ column density is a powerful tool for understanding the chemistry of planetary atmospheres. \ion{H}{1} Ly$\alpha$ and H$_{2}$ fluorescence are the brightest spectral features in the far-UV (FUV) spectrum of Jupiter (\citealt{Broadfoot79}), and are predicted to be the brightest features in the spectra of hot Jupiters (\citealt{Menager10,Menager13}).~\cite{Wolven97} demonstrated that the abundance of H$_2$ can be constrained by measuring the H$_2$ fluorescence in the heated (T $>$ 1000 K) atmosphere of Jupiter near the Shoemaker-Levy 9 (S-L9) comet impact site. In principle, these same techniques could be extended to exo-Jovian atmospheres (\citealt{Yelle04,France10}).

H$_2$ fluorescence can result from the photoexcitation of H$_2$ by Ly$\alpha$ emission from the host star. Observations of highly-active M dwarfs without known planetary systems,\footnote{Proxima Centauri has a recently discovered planet. See \citealt{Anglada16}} showed limited signs of \ion{O}{6}-pumped H$_2$ fluorescence, but the Ly$\alpha$-driven cascade was not observed (\citealt{Redfield02, France07}).~\cite{France13} (F13) noted the discovery of Ly$\alpha$-pumped H$_2$ features in the spectra of four M dwarf systems with confirmed planets. Due to the small number of targets and low signal-to-noise (S/N) of the F13 observations, they were unable to conclusively identify the origin of the emission, instead proposing three possible explanations:
\begin{enumerate}
\item {\it A planetary origin} - For Ly$\alpha$ fluorescence to occur in exoplanetary atmospheres, warm populations of H$_2$ are required. One possibility is that the H$_2$ is heated by impacts from rocky and icy planetismals, similar to the fluorescence observed at the S-L9 impact site~\citep{Wolven97}. Alternatively, the extreme ultraviolet (EUV; 100-912 \AA) emission from the M dwarf could heat portions of the exoplanet atmospheres to temperatures greater than 1500 K~\citep{Yelle04,Koskinen10,Loyd17}, generating conditions for Ly$\alpha$-excited H$_2$ fluorescence. Predictions for H$_2$O dissociation into H$_2$ . Both methods argue for the presence of a Jovian-mass planet, since they are comprised of large quantities of H$_2$. These planets are the most likely candidates for planetary H$_2$ fluorescence, but, as shown by F13, they are not necessary for the emission to exist. Another possible source could be a water-rich super-Earth that is undergoing rapid atmospheric mass loss. H$_2$ is a byproduct of the dissociation of H$_2$O by Ly$\alpha$ photons, and so the mass loss combined with the activity of the host star could create a favorable environment for fluorescence.~\cite{Slanger82} predict an H$_2$ production rate of 10\% when Ly$\alpha$ is the source of dissociation. This is in rough agreement with~\cite{Huebner79} who list a rate of $\sim$9\% for H$_2$O 1 AU from the Sun.

\item {\it A circumstellar disk origin} - Transit observations of short-period planets indicate that hydrodynamic mass loss may be ubiquitous on short-period planets (e.g.~\citealt{Vidal13,Kulow14,Ehrenreich15}). This gas could maintain a circumstellar envelope near the orbit of the planet~\citep{Haswell12,Fossati13}. Similar to the planetary origin hypothesis, the H$_2$ could come directly from Jovian-mass planets or from super-Earths via H$_2$O dissociation. F13 found that the line widths of the fluorescent features from GJ 436 were too narrow to originate near GJ 436b. In situations like this, the gas would have to have settled into a ``second-generation'' disk at a larger semi-major axis.

\item {\it A stellar origin} - Ly$\alpha$-pumped H$_2$ fluorescence has been detected in the spectra of sunspots~\citep{Jordan77} so it is plausible we are observing a similar process on M dwarfs. H$_2$ could also reside in the cooler region of the lower chromosphere that is predicted by current semi-empirical models~\citep{Fontenla16}. The measured radial velocities (RV) of H$_2$ in F13 were consistent with the stellar RV. If true, the lack of fluorescence seen in the previously observed highly-active M dwarfs may have been a result of high instrumental background levels.
\end{enumerate}

To determine the origin of the H$_2$ emission, we use observations from the ``Measurements of the Ultraviolet Spectral Characteristics of Low-mass Exoplanet Host Stars" (MUSCLES) {\it Hubble Space Telescope} ({\it HST}) Treasury Survey (\citealt{France16}) to build a catalog of H$_2$ emission features in 11 M dwarfs, with eight that are known host planets and three that are not. We measure the kinematics of the H$_2$ using spectra from the Cosmic Origins Spectrograph (COS) on {\it HST}. By comparing the results from the stars that host planets to those that do not, we determine whether the exoplanets are a plausible source of the H$_2$ emission. We also reanalyze the spectra of the four previously mentioned highly-active M dwarfs (flare stars; hereafter), using observations from the StarCAT catalog (\citealt{Ayres10}), to see if an H$_2$ signal may be present but obscured by the instrumental noise or atomic stellar emission lines. Combined the two samples (MUSCLES and flare stars) contain a total of 15 M dwarfs (nine that are known to host planets, six that are not).

In addition to the spectroscopic reconnaissance described above, we have undertaken a novel direct imaging experiment to constrain the spatial origin of fluorescent H$_{2}$. The FUV bandpass offers a potentially high planet/star contrast ratio for direct imaging due to the absence of photospheric emission and weak chromospheric continuum emission (see, e.g.,~\citealt{Loyd16}) from the host star at $\lambda$ $<$ 2000 \AA. For emission lines that are not found in the stellar atmosphere, the planet/star contrast ratio could, in principle, be greater than unity in the FUV. 

After determining that the H$_2$ fluorescence has a stellar origin (\S\ref{Discuss}), we verify that our hypothesis is physically reasonable by modeling the emitting region using a 1D radiative transfer code. The measured H$_2$ fluxes constrain the column density and temperature of the emitting gas. These results are compared to an M dwarf atmosphere model to further refine the spatial origin of the fluorescence. The model parameters are also compared to other astrophysical H$_2$ fluorescing regions, such as protoplanetary disks, to gain further insight into the structure of the emission.

This paper is structured as follows: \S\ref{ObsMorph} discusses the observing strategies and the methodology used when analyzing the three datasets utilized in the work: the MUSCLES Treasury Survey, the flare stars, and the {\it HST} direct imaging campaign. \S\ref{Sults} presents the results of the analysis, and \S\ref{Discuss} discusses the origin of the emission. \S\ref{model} describes the execution and results of the radiative transfer models, and \S\ref{conclude} summarizes our findings.

\section{Observations \& Methodology} \label{ObsMorph}
Our analysis focuses on two Ly$\alpha$-driven progressions: B-X (1-2)P(5) at 1216.07 \AA~and B-X (1-2)R(6) at 1215.73 \AA. Within this nomenclature, the B-X indicates that a molecule is excited from the electronic ground state (X) to a higher electronic state (B). Both progressions also involve ro-vibrational transitions, with the (1-2) indicating that the molecule moved from the $v''$ = 2 level in the electronic ground state to the $v'$ = 1 level in the excited state. Rotationally, both transitions have a $|\Delta J = 1|$, with P(5) showing that the rotational level decreases by 1, from $J''$ = 5 to $J'$ = 4 and R(6) showing that the rotational level increases by 1, from $J''$ = 6 to $J'$ = 7.  The [1,4] progression therefore refers to emission lines that cascade from the [$v',J'$] = [1,4] ro-vibrational level of the B electronic state, and the [1,7] progression refers to emission lines that cascade from the [$v',J'$] = [1,7] ro-vibrational level of the B electronic state.

These progressions are chosen because their central wavelengths overlap with the Ly$\alpha$ emission line. This fluorescence mechanism requires ro-vibrationally excited H$_2$ populations~\citep{Shull78} that are typically thermally excited in a warm medium. Once ``pumped" into an electronically excited state, the molecule cascades back down to the electronic ground state through one of several possible transitions. We spectroscopically observe this process by measuring the host of emission lines formed by this cascade of electron transitions from one of the two upper states.

\subsection{MUSCLES} \label{MUSmeth}
Observations of the planet-hosting MUSCLES stars were obtained using the COS G130M and G160M modes~\citep{Loyd16}, providing wavelength coverage of many important emission features, including, \ion{C}{2} $\lambda$1335 and \ion{C}{4} $\lambda$1548, as well as the brightest H$_2$ fluorescence lines. Ly$\alpha$ $\lambda$1216 was also in this bandpass, but was contaminated by geocoronal emission and required contemporaneous STIS observations to reconstruct its profile~\citep{Youngblood16}. The final science spectra were created following the procedure described by~\cite{Danforth16} where the spectra from the individual exposures were reduced using the CALCOS\footnote{http://www.stsci.edu/hst/cos/pipeline/CALCOSReleaseNotes/notes/} pipeline and then coadded together. The uncertainties of these coadded spectra were calculated using a low-count applicable Poisson uncertainty multiplied by a sensitivity function.

For the three MUSCLES stars without confirmed planets, radial velocity observations can rule out Msin($i$) $>$ $\sim$1 M$_{\rm Jup}$ planets within 2 AU~\citep{Bonfils13} and within $\sim$1 AU for GJ 1061 and HD 173739~\citep{Endl06,Bonfils13}. These three stars were observed using only the G160M mode. GJ 628 fell within this category, but has since been found to host at least three planets (\citealt{Wright16}). While having a single mode means these stars had fewer H$_2$ lines available, the wavelength range it covers includes the two brightest H$_2$ transitions that were used when comparing emission from planet hosts and non-hosts. These were (1-7)R(3) at 1489.63 \AA~and (1-7)P(5) at 1504.81 \AA. A description of each of the stars and their planetary systems can be found in Table~\ref{ovtable}. See~\cite{France16} for further details on the MUSCLES observing strategy.

To characterize the H$_2$ emission, features for each star were first identified by eye using the line list of~\cite{Lupu06}. The COS data was binned by pixel, so a seven-pixel wide boxcar function was used to smooth the data to match the size of the COS resolution element~\citep{Debes16}. Identified lines were then fitted using a Gaussian convolved with the COS line spread function (LSF) (\citealt{Kriss11,France12}). Due to the uncertainties in the COS wavelength solution (\citealt{Debes16}), the line centers of neighboring ions were used to correct the H$_2$ line centers. This was done by fitting a nearby emission line on either side of each fluorescence line and then linearly interpolating that feature's observed line center using the fitted and laboratory line centers. The choice to use a linear interpolation was motivated by~\cite{Linsky12}, who found a linear drift in their COS-measured velocities as a function of wavelength. These fits and corrected line centers were used to calculate the velocity centroids, Doppler broadening and H$_2$ emission line fluxes of the MUSCLES stars.

\begin{table}
\caption{Descriptions of MUSCLES and flare star samples
\label{ovtable}}
\centering
\begin{tabular}{ c c c c c c c }
\hline
Star & Spectral\,\tablenotemark{a} & d\,\tablenotemark{a} & log$_{10}$ L(Ly$\alpha$)\,\tablenotemark{b} & Exoplanet Mass\,\tablenotemark{c} & Semi-major Axis & Ref.\,\tablenotemark{d} \\
& Type & \lbrack pc\rbrack & \lbrack log$_{10}$ erg s$^{-1}$\rbrack & Msin($i$) \lbrack M$_{\oplus}]$ & \lbrack AU\rbrack & \\ \hline
\multicolumn{7}{c}{MUSCLES stars with confirmed planets} \\ \hline
GJ 176 & M2.5 & 9.4 & 27.61 & 8.3 & 0.066 &~\cite{Forveille09}\\
GJ 436 & M2.5 & 10.3 & 27.43 & 23.1 & 0.0287 & \cite{Maness07} \\
GJ 581 & M3 & 6.3 & 26.74 & 15.9, 5.3, & 0.041, 0.073, & \cite{Forveille11} \\
& & & & 6.0, 1.9 & 0.218, 0.029 & \\
GJ 628 & M4 & 4.3 & --- & 1.36, 4.25, & 0.036, 0.084, & \cite{Wright16} \\
& & & & 5.21 & 0.204 & \\
GJ 667C & M1.5 & 6.9 & 27.47 & 5.6, 4.2 & 0.050, 0.125 & \cite{Robertson14} \\
GJ 832 & M1 & 4.9 & 27.43 & 217.4, 5.3 & 3.6, 0.16 & \cite{Wittenmyer14} \\
GJ 876 & M4 & 4.7 & 27.01 & 619.7, 194.5, & 0.208, 0.130, & \cite{Rivera10} \\
& & & & 5.8, 12.5 & 0.021, 0.333 &\\
GJ 1214 & M6 & 13.0 & 26.63 & 6.4 & 0.0143 & \cite{Carter11} \\ \hline
\multicolumn{7}{c}{MUSCLES stars without confirmed planets} \\ \hline
GJ 887 & M2 & 3.3 & --- & --- & --- & \\
GJ 1061 & M5 & 3.7 & --- & --- & --- & \\
HD 173739 & M3 & 3.6 & --- & --- & --- & \\ \hline
\multicolumn{7}{c}{Flare Stars\,\tablenotemark{e}} \\ \hline
AD Leo & M3.5 & 4.7 & 28.42 & --- & --- & \\
AU Mic & M0 & 9.9 & 29.08 & --- & --- & \\
EV Lac & M3.5 & 5.1 & 27.93 & --- & --- & \\ 
Proxima Cen & M5.5 & 1.3 & 26.93 & 1.3 & 0.05 & \cite{Anglada16} \\ \hline
\multicolumn{7}{l}{$^{\rm a}$ Simbad spectral type and parallax distance} \\
\multicolumn{7}{l}{$^{\rm b}$ Calculated using fluxes from~\cite{Youngblood16} } \\
\multicolumn{7}{l}{$^{\rm c}$ Planet information is listed alphabetically by name} \\
\multicolumn{7}{l}{$^{\rm d}$ References for planetary system values} \\
\multicolumn{7}{l}{$^{\rm e}$ Data on all flare stars is from~\cite{Linsky13}} \\
\end{tabular}
\end{table}

\subsection{Flare Stars} \label{flares}
Spectra of the four flare stars were obtained from the StarCAT database~\citep{Ayres10}, which compiled all of the {\it HST}-Space Telescope Imaging Spectrograph (STIS) observations for a given star. All observations were performed using the 0.2$''$ x 0.2$''$ slit on the E140M mode of STIS. In all four cases, the coadded spectrum was used for analysis. A description of each of the stars can be found at the bottom of Table~\ref{ovtable}, and Table~\ref{stis_obs} lists the dates and exposure times of the spectra used in the coaddition. See~\cite{Ayres10} for a full description of the available datasets.

\begin{table}
\caption{Dates and Exposure times of STIS observations
\label{stis_obs}}
\centering
\begin{tabular}{ c c c }
\hline
Star & Date & Exp. Time \\
& [MJD] & [s] \\ \hline
AD Leo & 51614.151 & 13000 \\
& 51615.090 & 13000 \\
& 51616.096 & 13000 \\
& 51613.145 & 13000 \\
& 52426.297 & 10390 \\
& 52427.233 & 4630 \\ \hline
AU Mic & 51062.512 & 2130 \\ 
& 51062.570 & 2660 \\ 
& 51062.637 & 2660 \\ 
& 51062.704 & 2655 \\ \hline
EV Lac & 52172.698 & 1875 \\ 
& 52172.751 & 3015 \\ 
& 52172.818 & 3015 \\ 
& 52172.885 & 3015 \\ \hline
Proxima Cen & 51673.002 & 15120 \\
& 51672.040 & 20580 \\ \hline
\\
\end{tabular}
\end{table}

The StarCat catalog oversamples the STIS data and so the flare star spectra were smoothed to match the intrinsic resolution of the STIS data and then visually inspected for the (1-7)R(3) and (1-7)P(5) transition lines that are used for comparison between our different populations. Detected H$_2$ lines were analyzed using the procedure described in \S\ref{MUSmeth}, with the STIS LSF in place of the COS LSF. 

For flare stars that lacked signs of one or both of the H$_2$ emission lines, an upper limit was calculated. The expected location of the line was determined using nearby ions following the same procedure described in \S\ref{MUSmeth}. We calculated an effective integrated flux by multiplying each value in a $\pm$20 km s$^{-1}$ region around the line center by 10 km s$^{-1}$ (in wavelength space). This 10 km s$^{-1}$ width was chosen because it is the approximate size of a thermally broadened emission feature in a region where the warm H$_2$ could reside. This value also agrees well with the widths of the observed H$_2$ lines. The mean and standard deviation of these estimated flux limits were measured and used to calculate a 2-$\sigma$ upper limit on the H$_2$ flux.

\subsection{UV Imaging Observations of GJ 832}
Of the stars in the MUSCLES Treasury Survey, the GJ 832 system is unique in its proximity to Earth (4.95 pc) and the presence of a Jovian-mass planet at a relatively wide separation (3.6 AU) from the parent star~\citep{Bailey09,Wittenmyer14}. With apastron and periastron distances of 3.89 AU and 3.31 AU, the largest star-planet angular separations are between 0.79\arcsec\ and 0.68\arcsec\ over the course of an orbit. This allowed for angularly-resolved FUV images of the planetary environment in this system using the $\sim$0.1\arcsec\ angular resolution of the $HST$-Advanced Camera for Surveys/Solar Blind Channel (ACS/SBC). This data serves as a direct test of the hypotheses of an exoplanetary or circumstellar origin for the fluorescent emission.

We obtained ACS/SBC imaging in several FUV filters optimized to isolate the atomic and molecular emission of hydrogen. As the brightest expected UV features from a Jovian planet are Ly$\alpha$ and H$_2$, we obtained imaging observations using the F122M (Ly$\alpha$), F140LP (containing the majority of the H$_2$ lines), and F165LP (to control for the known red-leak in the ACS/SBC channel) filters. The observations were extended into two {\it HST} visits to include two spacecraft roll orientations in order to control for potential artifacts that could be misinterpreted as an exoplanetary signal. Observations were executed on 29 April 2016 and 08 June 2016 as part of the guest observing program 14100 (PI – K. France). Exposure times of 4700s, 2600s, and 2400s in F122M, F140LP, and F165LP, respectively, were used in each of the two visits for a total of eight spacecraft orbits. To minimize the geocoronal background, the F122M observations were obtained during orbital night (i.e., in Earth's shadow). Observations during orbital night reduce the Ly$\alpha$ background levels by a factor of $\sim$10 as the incidence of resonantly scattered solar photons is reduced. In the final F122M images, which had the largest Ly$\alpha$ geocoronal background, the star was approximately two orders of magnitude brighter than the background.

The ACS/SBC images were examined for signs of H$_2$ emission at the expected star-planet separation of 0.7\arcsec, which was calculated under the assumption that the planet was near its maximum separation (as viewed by the observer) during observation. This assumption was supported by the nearly circular orbit of GJ 832b (e = 0.08$\pm$0.09; ~\citealt{Wittenmyer14}) and the RV of GJ 832, which was near its minimum during observation (i.e. the star was approaching the observer at its maximum velocity, and hence the planet was receding at its maximum velocity). To ensure that the activity of GJ 832 did not drastically change during the various observations, we measured the stellar flux and compared it to the spectroscopically measured fluxes from the MUSCLES Treasury program. A full description of this process can be found in Appendix~\ref{obs_ap}. TinyTim point-spread-function (PSF) models were downloaded from the STScI website\footnote{\tt http://tinytim.stsci.edu/cgi-bin/tinytimweb.cgi} for the appropriate camera+filter combination and location of the star in the images. We compared the angular extent of the signal from GJ 832 to the TinyTim PSFs in order to explore the wings of the stellar flux for excess emission from GJ 832b (\S\ref{dir_obs}).

\section{Results} \label{Sults}
\subsection{Spectral Analysis} \label{M-sults}
Figure~\ref{spec} shows a subsection of the MUSCLES and flare star spectra. Table~\ref{h2num} lists the number of H$_2$ fluorescence lines observed for each star. Table~\ref{h2num} only includes lines that are used in the spectroscopic analysis. It is immediately clear that both MUSCLES stars with and without known planets show signs of H$_2$ fluorescence. Only one MUSCLES star (GJ 1214) lacks any H$_2$ emission features, most likely due to the low S/N of the observations, and so it is excluded from our kinematic analysis. For the flare stares, where we only searched for the two brightest transitions, both (1-7)R(3) and (1-7)P(5) are found in the spectra of AD Leo and Proxima Centauri. (1-7)P(5) is also observed in AU Mic, but the noise level around (1-7)R(3) means that a confident identification of the feature cannot be made.

\begin{figure}
\centering
\includegraphics[width=150mm,keepaspectratio]{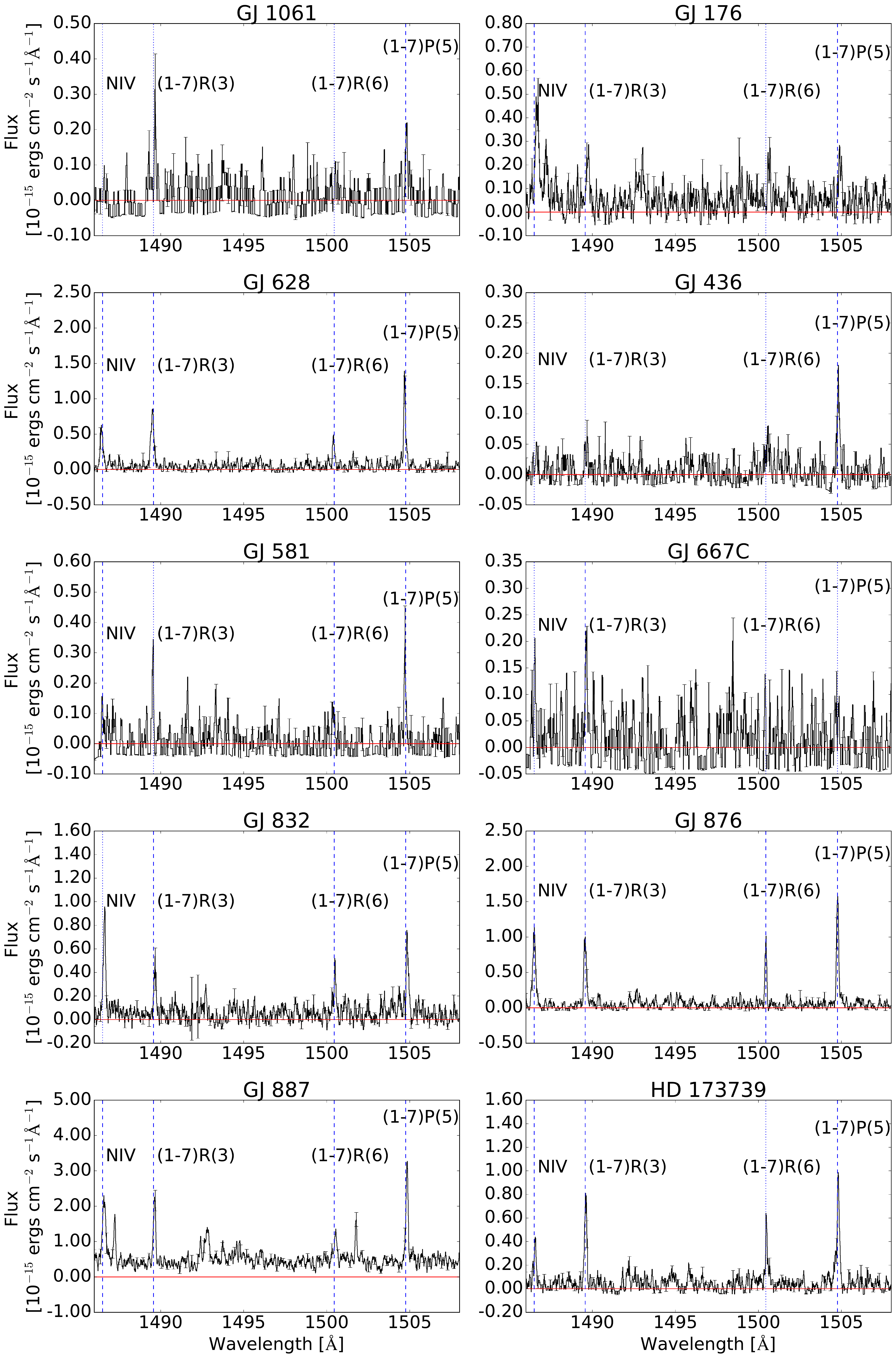}
\caption{Spectra of the MUSCLES and flare stars. The y = 0 line is included in red, for reference. A blue vertical line has been plotted at the laboratory wavelength of the lines of interest. Dashed lines indicate features that were included in the analysis, while dotted lines indicate emission features that may be identifiable by eye (e.g. (1-7)R(3) for GJ 1061) but were not included in the analysis for one of a variety of reasons. These include unresolved emission line profiles, contamination by other spectral features, or a lack of available ions for correcting the H$_2$ line centers.}
\label{spec}
\end{figure}

\begin{figure} \ContinuedFloat
\centering
\includegraphics[width=150mm,keepaspectratio]{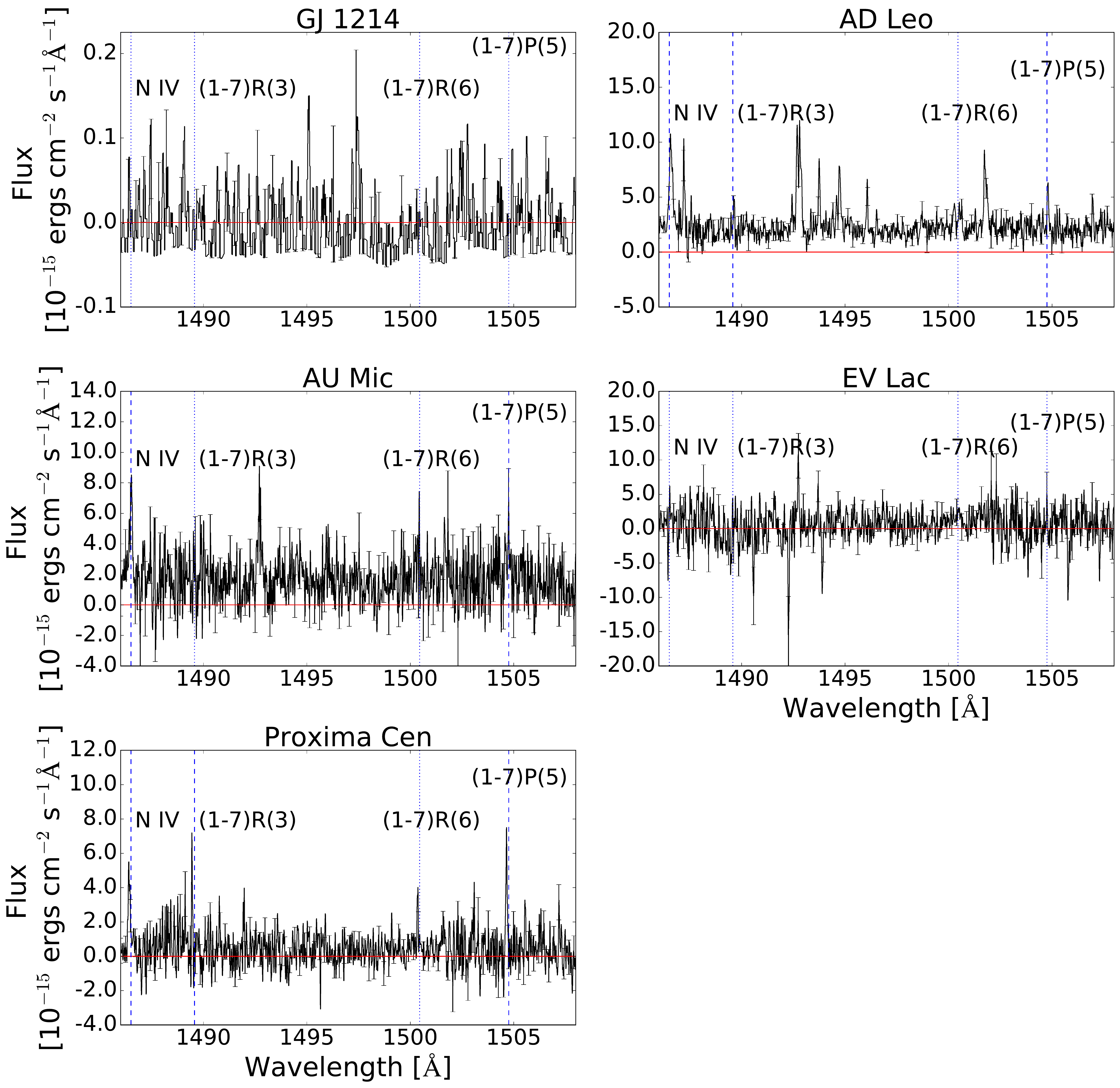}
\caption{continued.}
\end{figure}

\begin{table}
\caption{Number of identified fluorescent lines from each progression
\label{h2num}}
\centering
\begin{tabular}{ c c c }
\hline
Star & [1,4] & [1,7] \\ \hline
GJ 176 & 8 & 7 \\
GJ 436 & 4 & 6 \\
GJ 581 & 6 & 7 \\
GJ 628 & 7 & 5 \\
GJ 667C & 5 & 6 \\
GJ 832 & 9 & 8 \\
GJ 876 & 11 & 10 \\
GJ 1214 & 0 & 0 \\ \hline
GJ 887 & 6 & 5 \\
GJ 1061 & 2 & 3 \\
HD 173739 & 5 & 5 \\ \hline
AD Leo & 2 & 0 \\ 
AU Mic & 1 & 0 \\
EV Lac & 0 & 0 \\
Proxima Cen & 2 & 0 \\ \hline \\
\end{tabular}
\end{table}

\subsubsection{Radial Velocities} \label{Mrv}
\begin{figure}
\centering
\includegraphics[width=\textwidth]{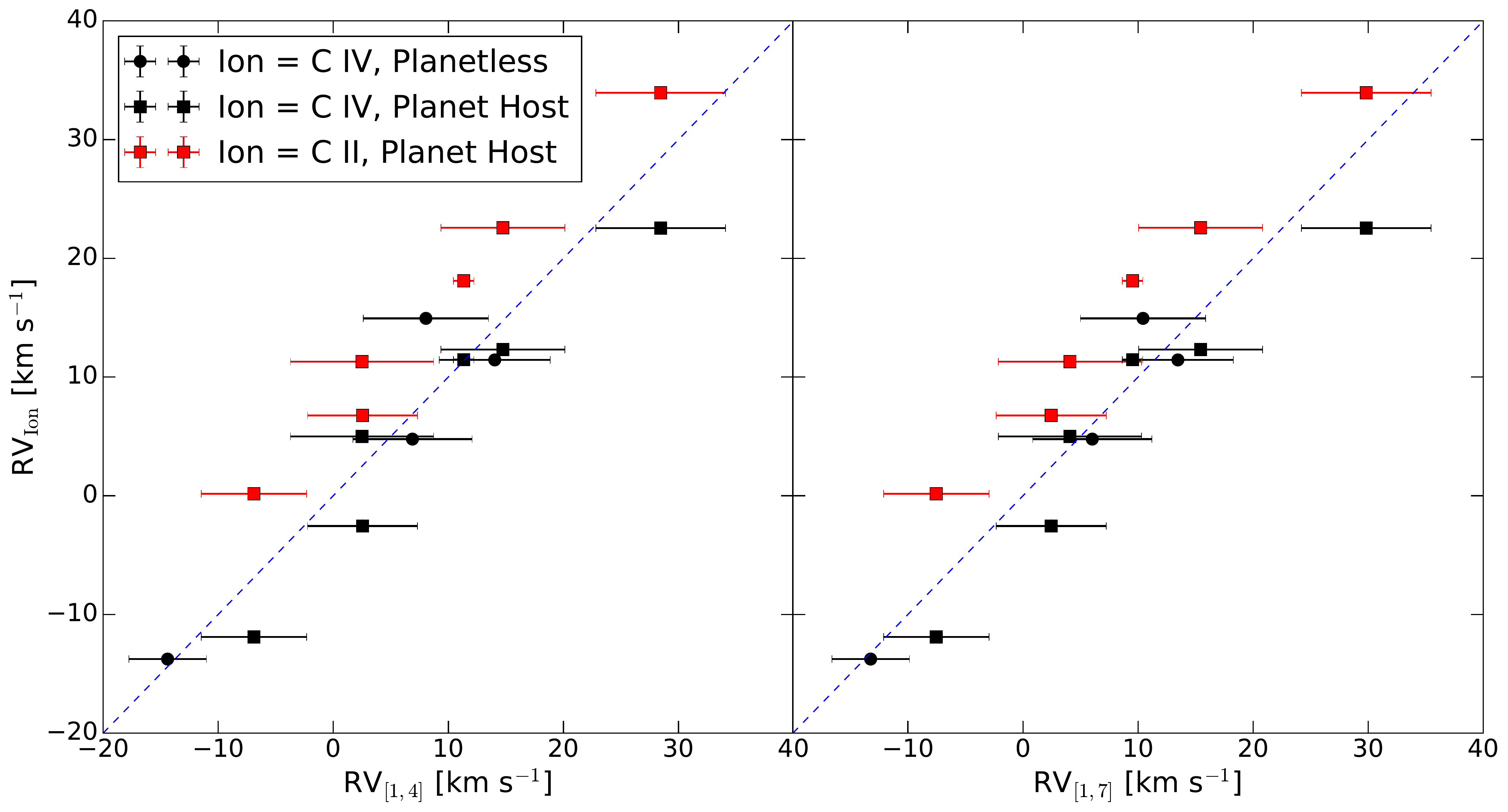}
\caption{The radial velocities calculated using the average of the transitions from the [1,4] (left) and [1,7] (right) progressions, compared to the radial velocities from \ion{C}{4} (black points) and \ion{C}{2} (red points). The square points are the planet-hosting MUSCLES stars, while the circles are the MUSCLES stars with no known planets. The 1:1 line has been included, for reference. Errors in the \ion{C}{4} and \ion{C}{2} RVs are included, but fall within the size of the data point. The errors do not include the 15 km s$^{-1}$ COS wavelength solution uncertainty.}
\label{rvd}
\end{figure}

Average RVs are calculated using the corrected line centers of the observed fluorescent transitions (i.e., the lines included in Table~\ref{h2num}) from each of the two progressions. We also measure the velocity centroids of two stellar ionic species: \ion{C}{2} at 1335.71 \AA~and \ion{C}{4} at 1548.20 \AA, to compare with published RV measurements (Figure~\ref{rvd} and Table~\ref{RV}). The \ion{C}{2} line is outside of the wavelength range of the G160M COS mode, so stars that are not confirmed planet hosts have no \ion{C}{2} RV.

The RVs calculated using \ion{C}{2} are, on average, larger than those measured using \ion{C}{4} by 9.0 $\pm$ 2.2 km s$^{-1}$. We believe that this is a result of the differences in the wavelength solutions between the G130M and G160M modes of COS. For GJ 1061, the calculated \ion{C}{4} RV is inconsistent with the published -20 km s$^{-1}$~\citep{Rodgers74}. To investigate this discrepancy, we independently calculate the RV of GJ 1061 using data from the High Accuracy Radial Velocity Planet Search (HARPS). Using a selection of spectral features, we measure a value of -14.30 $\pm$ 0.29 km s$^{-1}$, which is still a large difference but is within the COS accuracy.

\begin{table}
\caption{Velocity Centroids \lbrack km s$^{-1}$\rbrack~and FWHMs \lbrack km s$^{-1}$\rbrack~of MUSCLES stars
\label{RV}}
\centering
\begin{tabular}{ c c c c c c c c c }
\hline
Star & Literature & \ion{C}{4} & \ion{C}{2} & [1,4]\,\tablenotemark{a} & [1,7]\,\tablenotemark{a} & FWHM$_{\rm [1,4]}$ & FWHM$_{\rm [1,7]}$ & Ref.\,\tablenotemark{b} \\ \hline
GJ 176 & 26.4 $\pm$ $<$0.1 & 22.6 $\pm$ 0.2 & 32.0 $\pm$ 0.2 & 28.5 $\pm$ 5.6 & 29.8 $\pm$ 4.0 & 23.7 $\pm$ 9.4 & 11.1 $\pm$ 0.7 & 1 \\
GJ 436 & 9.6 $\pm$ 0.1 & 11.5 $\pm$ 0.2 & 18.1 $\pm$ 0.1 & 11.3 $\pm$ 0.9 & 9.5 $\pm$ 5.7 & 25.1 $\pm$ 11.4 & 22.7 $\pm$ 4.2 & 1 \\
GJ 581 & -9.2 $\pm$ $<$0.1 & -11.9 $\pm$ 0.2 & 0.2 $\pm$ 0.1 & -6.9 $\pm$ 4.6 & -7.5 $\pm$ 3.4 & 14.7 $\pm$ 6.4 & 17.3 $\pm$ 9.4 & 1\\
GJ 628 & -21.2 $\pm$ 0.1 & -13.8 $\pm$ 0.1 & --- & -14.4 $\pm$ 3.4 & -13.2 $\pm$ 2.5 & 22.2 $\pm$ 10.9 & 24.5 $\pm$ 11.2 & 1 \\
GJ 667C & 6.4 $\pm$ $<$0.1 & 5.0 $\pm$ 0.2 & 11.3 $\pm$ 0.2 & 2.5 $\pm$ 6.2 & 4.1 $\pm$ 2.7 & 17.1 $\pm$ 7.8 & 13.4 $\pm$ 4.1 & 1\\
GJ 832 & 4.3 $\pm$ 1.8 & 12.3 $\pm$ 0.1 & 22.6 $\pm$ 0.1 & 14.7 $\pm$ 5.4 & 15.4 $\pm$ 4.1 & 19.9 $\pm$ 4.5 & 24.3 $\pm$ 6.8 & 2\\
GJ 876 & -1.6 $\pm$ 0.2 & -2.6 $\pm$ 0.1 & 6.8 $\pm$ 0.1 & 2.5 $\pm$ 4.8 & 2.5 $\pm$ 5.5 & 22.2 $\pm$ 7.6 & 19.6 $\pm$ 7.4 & 1\\ \hline
GJ 887 & 7.5 $\pm$ 0.5 & 11.4 $\pm$ $<$0.1 & --- & 14.0 $\pm$ 4.8 & 13.5 $\pm$ 5.3 & 21.5 $\pm$ 8.4 & 22.5 $\pm$ 5.1 & 3\\
GJ 1061 & -14.3 $\pm$ 0.3 & 14.9 $\pm$ 0.2 & --- & 8.1 $\pm$ 5.4 & 10.4 $\pm$ 3.4 & 17.3 $\pm$ 2.0 & 22.2 $\pm$ 17.2 & 4 \\
HD 173739 & -1.1 $\pm$ 0.1 & 4.8 $\pm$ 0.1 & --- & 6.9 $\pm$ 5.2 & 6.0 $\pm$ 5.1 & 24.1 $\pm$ 8.8 & 17.2 $\pm$ 7.1 & 1 \\ \hline
\multicolumn{7}{l}{{\bf Note: } Above errors do not include the 15 km s$^{-1}$ COS wavelength solution uncertainty.} \\
\multicolumn{7}{l}{$^{\rm a}$ Average of all the lines that fluoresce from the designated upper state.} \\
\multicolumn{7}{p{0.81\textwidth}}{$^{\rm b}$ References for literature RV: 1)~\cite{Nidever02}; 2)~\cite{Gontcharov06}; ~\cite{Desidera15}; 4) See \S\ref{Mrv}} \\
\\
\end{tabular}
\end{table}

\subsubsection{Doppler Broadening}
To quantify the Doppler broadening, average FWHMs are measured for the two progressions using the observed fluorescent transitions (Table~\ref{RV}). A FWHM was also calculated for each MUSCLES star with observed H$_2$ emission by coadding all of the lines listed in Table~\ref{h2num} and fitting the result. Those measurements agreed with the average values and are not included here. Only GJ 176 has inconsistent FWHMs between the two progressions, and we expect that this is the result of low S/N leading to poorly sampled line profiles within the data (see, e.g., Figure~\ref{planet_em}, upper left).

\subsubsection{Fluxes}
\begin{figure}
\centering
\includegraphics[width=0.5\textwidth]{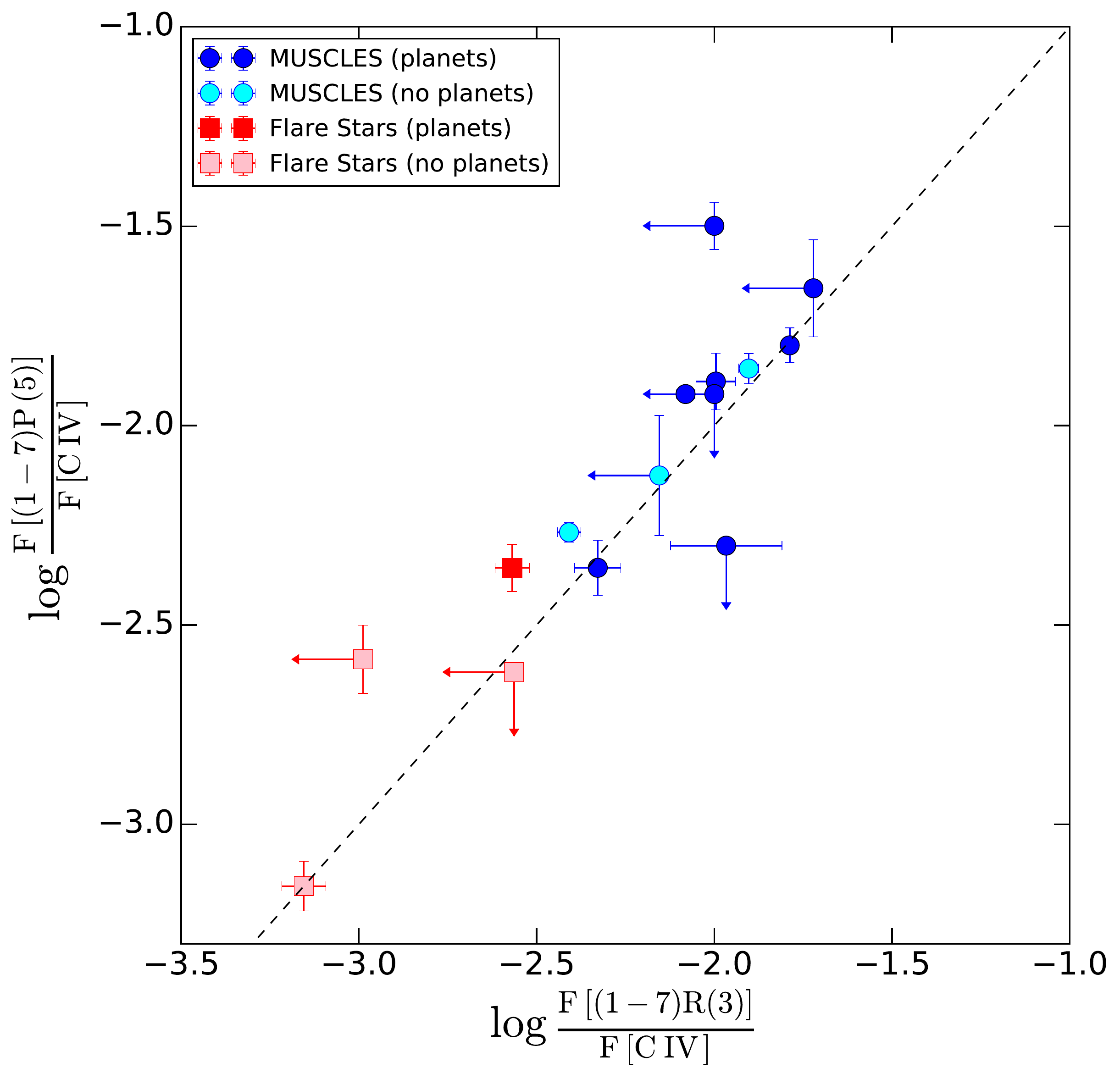}
\caption{Flux ratios using the two brightest H$_2$ fluorescence lines, (1-7)R(3) and (1-7)P(5), and the \ion{C}{4} lines. The plot includes ratios for the MUSCLES stars and flare stars (from Table~\ref{fluxes}). The 1:1 line has been included, for reference. Limits are designated by arrows in place of error bars. No difference is seen between the populations of planet hosts and non-hosts.}
\label{ratio_plot}
\end{figure}

The amount of H$_2$ emission varies from star to star, so to facilitate comparisons, we normalize the fluxes by calculating the ratio of the H$_2$ flux to stellar ion flux. The lines used for this calculation are: (1-7)R(3) at 1489.63 \AA, (1-7)P(5) at 1504.81 \AA, \ion{C}{2} at 1335.71 \AA, and the sum of the two \ion{C}{4} lines at 1548.20 \AA~and 1550.77 \AA. The \ion{C}{2} line at 1334.53 \AA~is not included because it is attenuated by the ISM (\citealt{Youngblood16}). These lines were chosen because they are the most prominent lines in the stellar spectra of M dwarfs in the COS G130M (for \ion{C}{2}) and G160M (for \ion{C}{4}) bandpasses and they have significantly different formation temperatures, providing redundancy in the event that an individual star in the sample has a drastically different atmospheric heating profile. The measured fluxes, flux upper limits (see \S\ref{flares}), and corresponding H$_2$-to-carbon flux ratios are listed in Table~\ref{fluxes}; the ratios are also shown in Figure~\ref{ratio_plot}.

\begin{table}[h]
\caption{Fluxes and flux ratios of MUSCLES and flare stars
\label{fluxes}}
\centering
\begin{tabular}{ c | c c | c c | c c c c }
\hline
\rule{0pt}{4ex} Star & \ion{C}{4}\,\tablenotemark{a} & \ion{C}{2} & (1-7)R(3) & (1-7)P(5) & $\frac{\mbox{(1-7)R(3)}}{\mbox{\ion{C}{4}}}$ & $\frac{\mbox{(1-7)R(3)}}{\mbox{\ion{C}{2}}}$
& $\frac{\mbox{(1-7)P(5)}}{\mbox{\ion{C}{4}}}$ & $\frac{\mbox{(1-7)P(5)}}{\mbox{\ion{C}{2}}}$ \\[1.5ex]
& \multicolumn{2}{c |}{[10$^{-15}$ erg cm$^{-2}$ s$^{-1}$]} & \multicolumn{2}{c |}{[10$^{-17}$ erg cm$^{-2}$ s$^{-1}$]} & \multicolumn{4}{c}{[10$^{-2}$]} \\ \hline
GJ 176 & 12.0 $\pm$ 0.3 & 5.5 $\pm$ 0.4 & 5.6 $\pm$ 0.8 & 5.3 $\pm$ 0.8 & 0.5 $\pm$ 0.1 & 1.0 $\pm$ 0.2 & 0.4 $\pm$ 0.1 & 1.0 $\pm$ 0.2 \\
GJ 436 & 1.1 $\pm$ 0.1 & 1.2 $\pm$ 0.1 & $<$1.2 & 3.6 $\pm$ 0.4 & $<$1.0 & $<$1.0 & 3.2 $\pm$ 0.4 & 3.1 $\pm$ 0.4 \\
GJ 581 & 2.0 $\pm$ 0.2 & 0.5 $\pm$ $<$0.1 & $<$3.7 & 4.4 $\pm$ 1.2 & $<$1.9 & $<$1.7 & 2.2 $\pm$ 0.6 & 8.5 $\pm$ 2.4 \\ 
GJ 628 & 12.2 $\pm$ 0.3 & --- & 19.8 $\pm$ 0.9 & 19.3 $\pm$ 1.8 & 1.6 $\pm$ 0.1 & --- & 1.6 $\pm$ 0.2 & --- \\
GJ 667C & 3.0 $\pm$ 0.3 & 0.7 $\pm$ 0.1 & 3.3 $\pm$ 1.2 & $<$1.4 & 1.1 $\pm$ 0.4 & 4.9 $\pm$ 1.8 & $<$0.5 & $<$2.0 \\ 
GJ 832 & 8.1 $\pm$ 0.2 & 3.8 $\pm$ 0.1 & 8.3 $\pm$ 1.0 & 10.5 $\pm$ 1.7 & 1.0 $\pm$ 0.1 & 2.2 $\pm$ 0.3 & 1.3 $\pm$ 0.2 & 2.8 $\pm$ 0.4 \\
GJ 876 & 23.7 $\pm$ 0.5 & 10.7 $\pm$ 0.5 & 19.7 $\pm$ 1.1 & 28.3 $\pm$ 1.1 & 0.8 $\pm$ 0.1 & 1.8 $\pm$ 0.1 & 1.2 $\pm$ $<$0.1 & 2.6 $\pm$ 0.2 \\ 
GJ 1214 & 0.5 $\pm$ 0.1 & 0.1 $\pm$ $<$0.1 & $<$0.6 & $<$0.6 & $<$1.1 & $<$6.8 & $<$1.2 & $<$7.9 \\ \hline
Average\,\tablenotemark{b} & --- & --- & --- & --- & 1.0 $\pm$ 0.4 & 2.5 $\pm$ 1.7 & 1.7 $\pm$ 0.9 & 3.6 $\pm$ 2.8 \\ \hline \hline
GJ 887\,\tablenotemark{c} & 91.4 $\pm$ 1.0 & --- & 36.1 $\pm$ 2.3 & 49.7 $\pm$ 3.1 & 0.4 $\pm$ $<$0.1 & --- & 0.5 $\pm$ $<$0.1 & --- \\
GJ 1061 & 4.5 $\pm$ 0.2 & --- & $<$3.0 & 3.4 $\pm$ 1.2 & $<$0.7 & --- & 0.8 $\pm$ 0.3 & --- \\
HD 173739 & 11.8 $\pm$ 0.3 & --- & 14.7 $\pm$ 0.9 & 16.4 $\pm$ 1.3 & 1.3 $\pm$ 0.1 & --- & 1.4 $\pm$ 0.1 & --- \\ \hline
Average & --- & --- & --- & --- & 0.8 $\pm$ 0.6 & --- & 0.9 $\pm$ 0.4 & --- \\ \hline \hline
AD Leo & 500.9 $\pm$ 3.1 & 148.9 $\pm$ 9.0 & 25.5 $\pm$ 5.3 & 36.7 $\pm$ 6.5 & 0.1 $\pm$ $<$0.1 & 0.2 $\pm$ $<$0.1 & 0.1 $\pm$ $<$0.1 & 0.3 $\pm$ 0.1\\
AU Mic & 305.0 $\pm$ 3.4 & 126.5 $\pm$ 6.1 & $<$632.3 & 79.1 $\pm$ 15.5 & $<$0.1 & $<$0.2 & 0.3 $\pm$ 0.1 & 0.6 $\pm$ 0.1\\
EV Lac & 126.1 $\pm$ 3.0 & 36.5 $\pm$ 1.5 & $<$34.5 & $<$30.4 & $<$0.3 & $<$0.9 & $<$0.2 & $<$0.8 \\
Proxima Cen & 129.5 $\pm$ 1.0 & 24.5 $\pm$ 1.1 & 0.4 $\pm$ $<$0.1 & 0.6 $\pm$ 0.1 & 0.3 $\pm$ $<$0.1 & 1.4 $\pm$ 0.2 & 0.4 $\pm$ 0.1 & 2.3 $\pm$ 0.3\\ \hline
\multicolumn{9}{l}{{\bf Note:} Listed errors are 1-$\sigma$ uncertainties. Fluxes with no trailing errors are 3-$\sigma$ upper limits.}\\
\multicolumn{9}{l}{$^{\rm a}$ Sum of the 1548.20 and 1550.77 \AA~lines. } \\
\multicolumn{9}{l}{$^{\rm b}$ Does not include GJ 1214.} \\
\multicolumn{9}{l}{$^{\rm c}$ A \ion{C}{4} flare occurred during observations of GJ 887, leading to significantly larger \ion{C}{4} fluxes.} \\
\end{tabular}
\end{table}

\subsection{Direct Imaging Observations} \label{dir_obs}
\begin{figure}
\centering
\begin{tabular}{c c}
\includegraphics[width=66.7mm]{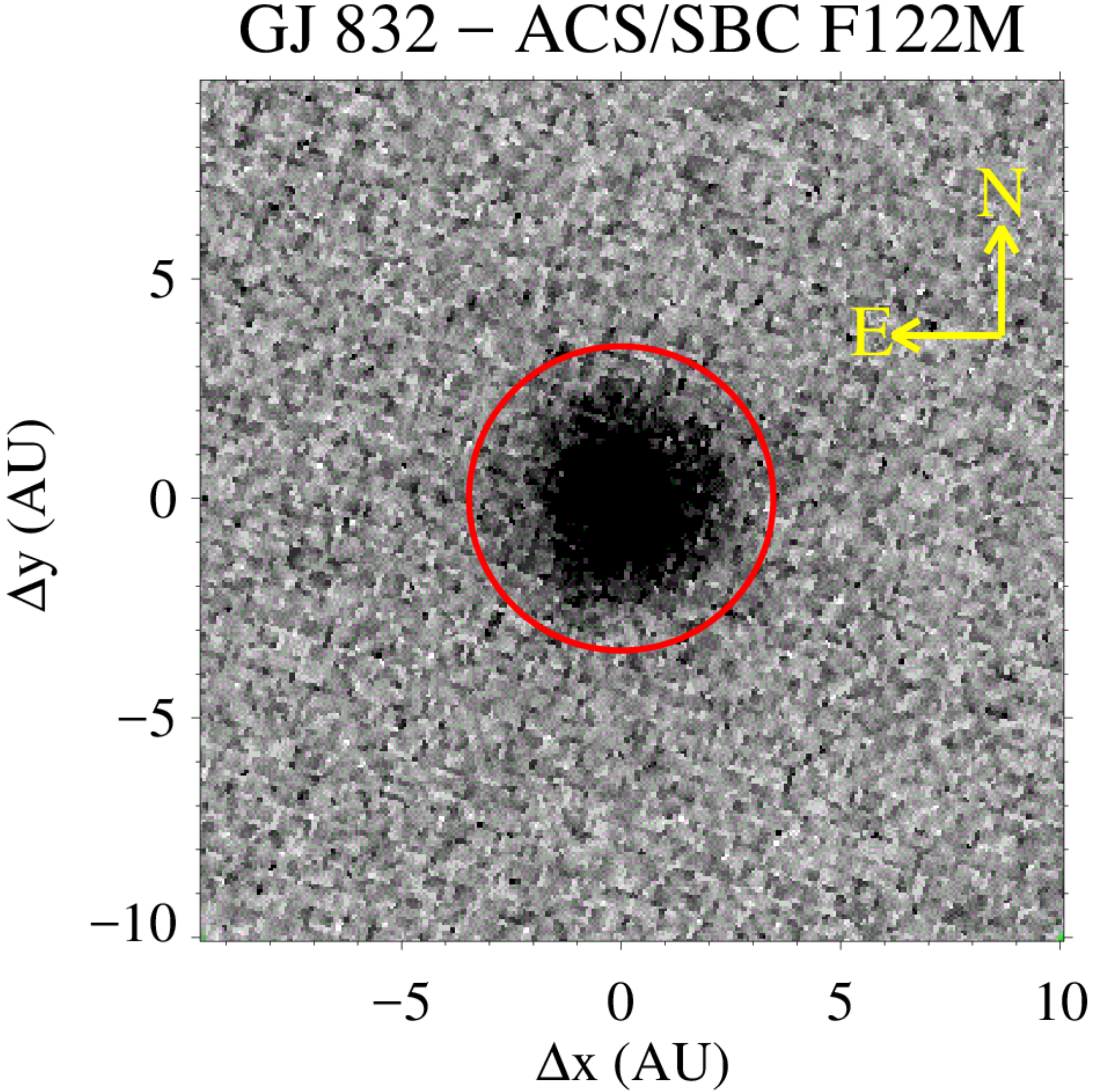} & \includegraphics[width=63.8mm]{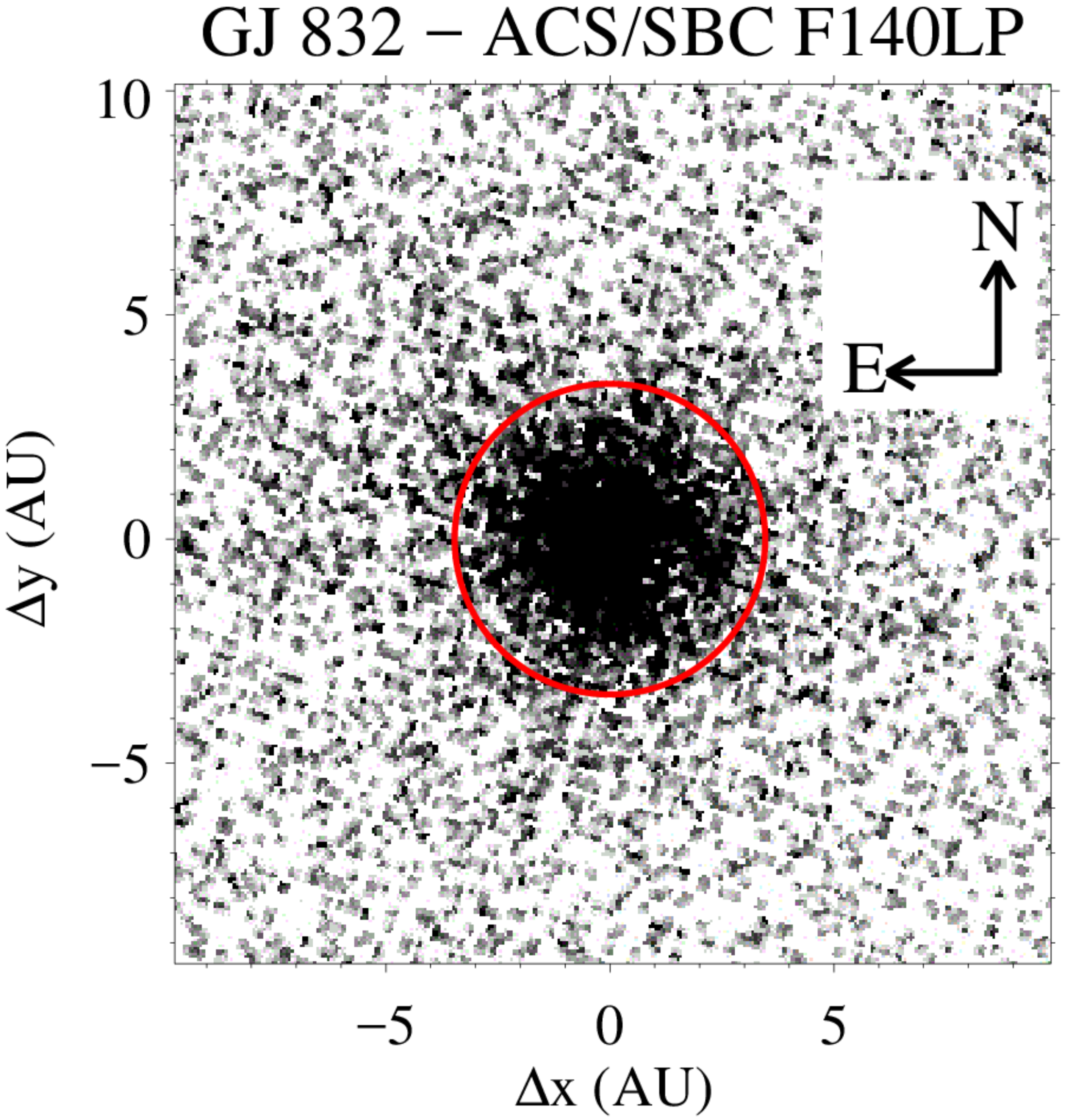}\\
\multicolumn{2}{c}{\includegraphics[width=63.8mm]{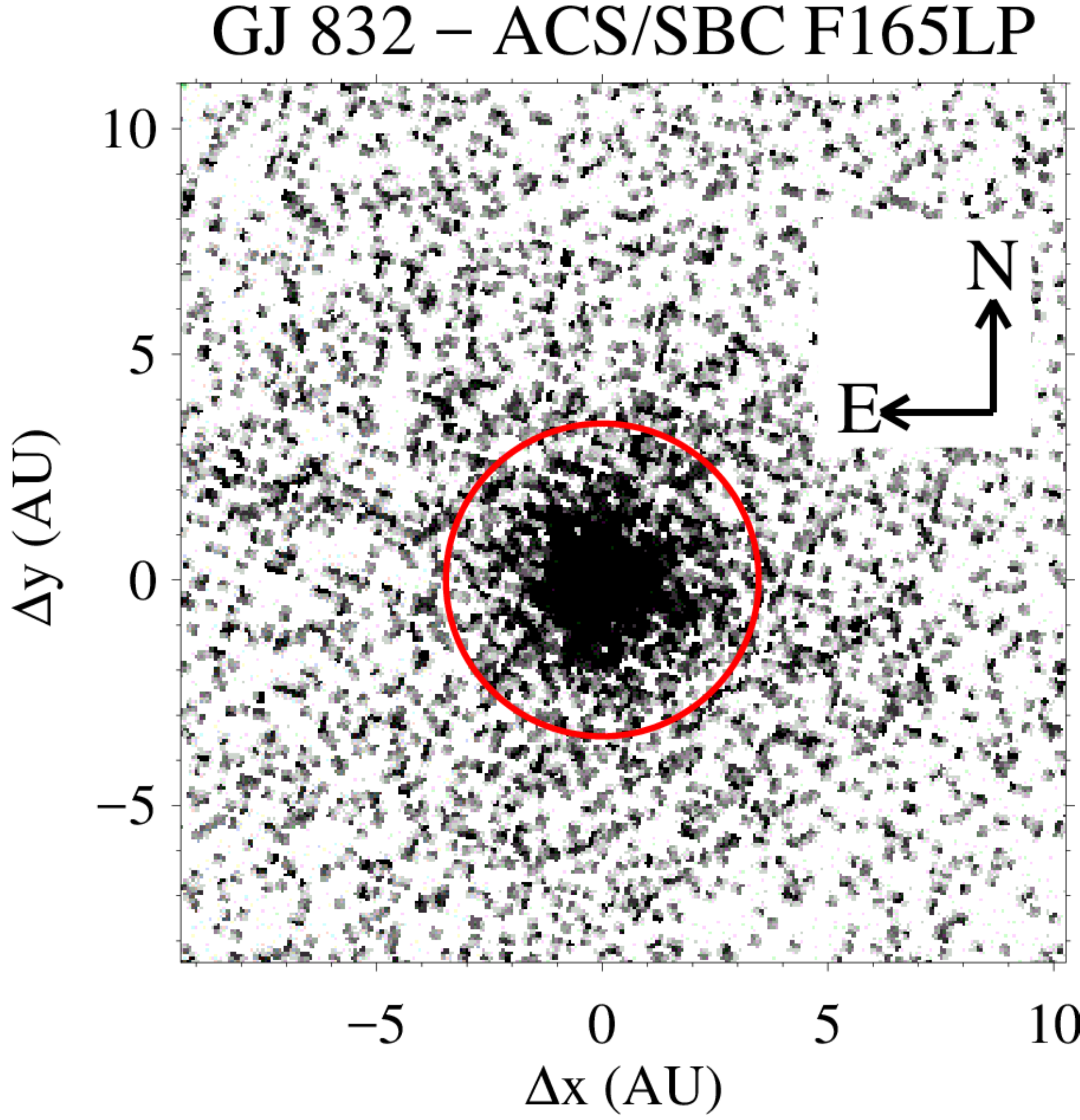} }\\
\end{tabular}
\caption{
\label{cosovly} $HST$ ACS/SBC images of the GJ 832 system in the F122M, F140LP, and F165LP filters. The orientation is rotated north up and the spatial scale is shown for each image. The expected $\sim$0.7\arcsec\ location of GJ 832b is shown as a solid red circle; no emission at the planetary position is detected above the stellar PSF. The location of the system is (ep=J2000) RA = 21:33:33.98, Dec = -49:00:32.42 and it has proper motions in RA of -46.05 mas yr$^{-1}$ and in Dec of -817.63 mas yr$^{-1}$~\citep{van07}. Upper limits on the emission from GJ 832b are presented in \S\ref{dir_obs}.
}
\end{figure}

\begin{figure}
\centering
\includegraphics[width=125mm]{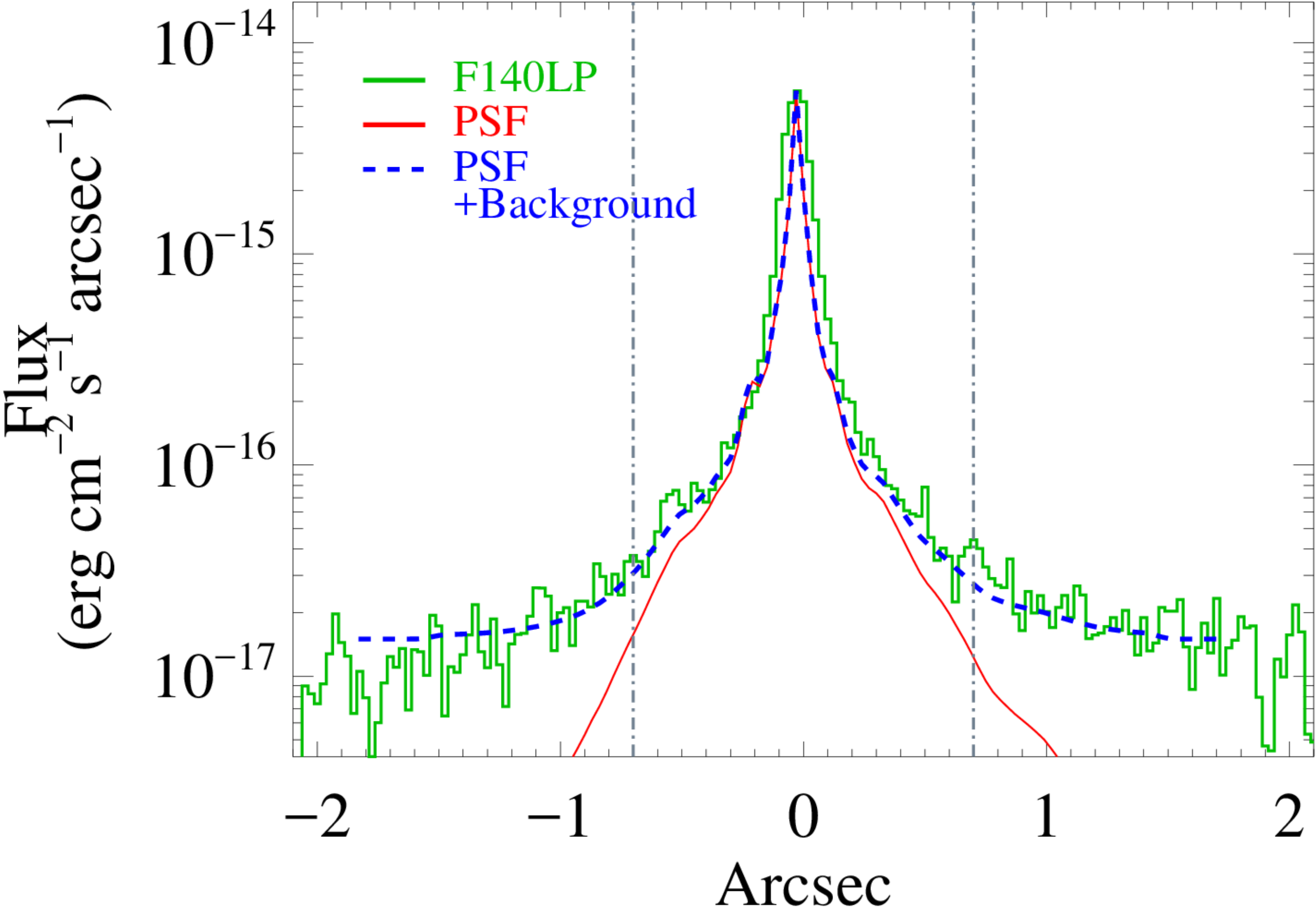}
\caption{
\label{cosovly_F140} Azimuthally-averaged angular profile of the ACS/SBC F140LP observations of GJ 832 (green histogram). The TinyTim F140LP PSF is shown in red and the blue dashed line is the PSF with a constant background noise level of $B_{140}$~=~1.5 $\times$~10$^{-17}$ erg cm$^{-2}$ s$^{-1}$ arcsec$^{-1}$ added. The expected location of GJ 832b in this representation is at $\sim$0.7\arcsec, shown as the dash-dot gray lines. 
}
\end{figure}

Images of all three ACS/SBC bands are shown in Figure~\ref{cosovly}. The expected 0.7\arcsec\ star-planet separation is shown as a red circle. No exoplanetary or circumstellar signal is detected in the ACS/SBC imaging observations of GJ 832, and the stellar photometry measured in the three bands is consistent between visits as well as with the spectroscopically measured fluxes from the MUSCLES Treasury Survey (Appendix~\ref{obs_ap}). The collapsed one-dimensional angular profile of the star in the F140LP imaging band is shown in Figure~\ref{cosovly_F140}. The data in the F140LP and F165LP bands are consistent with a point source plus constant background contribution, $B_{\rm filter}$, with $B_{140}$~=~1.5 $\times$~10$^{-17}$ erg cm$^{-2}$ s$^{-1}$ arcsec$^{-1}$ and $B_{165}$~=~4.0~$\times$~10$^{-17}$ erg cm$^{-2}$ s$^{-1}$ arcsec$^{-1}$ for F140LP and F165LP, respectively. 

The broad PSF of the ACS/SBC filters and the small working angle set by the star-planet separation complicate the determination of a photometric upper limit for GJ 832b. The exact orbital geometry of the planet during observation is not known, only that it was near the maximum star-planet separation. This places the planet at $\sim$0.7\arcsec~from the star in any orbital geometry. To measure the FUV flux upper limit, we average the angular profile of each image, centered on the expected radial distance of the planet, over the range 0.65\arcsec-0.75\arcsec~from the peak stellar flux. We place 3-$\sigma$ flux upper limits on the FUV emission from GJ 832b of 7.5~$\times$~10$^{-16}$ erg cm$^{-2}$ s$^{-1}$ in F122M, 4.9~$\times$~10$^{-17}$ erg cm$^{-2}$ s$^{-1}$ in F140LP, and 1.2~$\times$~10$^{-16}$ erg cm$^{-2}$ s$^{-1}$ in F165LP. 

\section{Discussion} \label{Discuss}
Using the results described in \S\ref{Sults}, we evaluate the three origin hypotheses to determine which one is best described by the observations.
\subsection{Planetary origin} \label{po}
If the H$_2$ fluorescence has a planetary origin, we would expect to see the emission in the MUSCLES stars with confirmed planets but not in those without. In this scenario, the emission features could be shifted relative to the line centers of the stellar features, depending on the RV of each planet during observations. These requirements come with a number of caveats that need to be addressed to fully rule out a planetary origin. Specifically, H$_2$ emission could be observed simultaneously from the star and planet, since both may have regions where H$_2$ fluorescence could occur. Additionally, the phase of the planet during observation could mean that its emission features are spectrally unresolvable from those of the host star.

H$_2$ fluorescence is observed in 10 out of the 11 MUSCLES stars (Table~\ref{h2num}) in our sample, including those without known planets, in disagreement with the predictions of the planetary origin hypothesis. GJ 1214 is the only star without detected H$_2$ emission features, but the limits of the flux ratios (Table~\ref{fluxes}) indicate that the noise floor may be too high to detect a typical ratio for this star. The RVs listed in Table~\ref{RV} show that all of the H$_2$ line centers are consistent with the published RVs within the accuracy of the COS wavelength solution. This does not rule out a planetary origin but does require that all of the observations occurred when the RV of each planet was consistent with that of its host star. This is discussed in more detail in Section~\ref{PlanRV}.

These spectroscopic results are further supported by our direct imaging campaign, where we do not observe any planetary emission at the expected star-planet separation. Using the observed flux in the {\it HST}-COS spectrum that falls within the F140LP bandpass (Appendix~\ref{obs_ap}, F$_{\rm COS}$(140LP) = 3.8 x 10$^{-14}$ erg cm$^{-2}$ s$^{-1}$) and the fractional contribution of H$_2$ to that emission (Appendix~\ref{obs_ap}, f$_{\rm H_2}$ = 14.6\%), we estimate a spectroscopically-observed H$_2$ flux in the {\it HST}-COS data of 5.5 x 10$^{-15}$ erg cm$^{-2}$ s$^{-1}$. This is two orders of magnitude larger than the 3-$\sigma$ upper limit on the FUV flux in the F140LP band of GJ 832b (\S\ref{dir_obs}), ruling out a planetary signal as the sole source of the H$_2$ fluorescence emission.

\subsubsection{Planetary contribution to the spectrum} \label{PlanRV}
While the above results rule out planetary emission as the only source of the H$_2$ fluorescence, it is still possible that the planets are contributing some signal to the observations. If this were the case, we would expect the confirmed hosts to have a consistently larger H$_2$ fluorescence flux as a result of the additional planetary signal. We can look for this signal by comparing the relative fluxes between the confirmed and unconfirmed hosts. Figure~\ref{ratio_plot} shows that the measured H$_2$-to-stellar ion flux ratios between the two populations are consistent. This comparison assumes that every planet has an RV consistent with its host star, which may not always be the case. Planets with potentially resolvable H$_2$ emission features can be identified by reconstructing their phases during observations.

Both of the H$_2$ lines used to calculate the flux ratios are located in the COS G160M bandpass, so we estimate the phase of each planet during these observations using the orbit parameters provided by the references in Table~\ref{ovtable}. The RVs of the planets in the rest frame of the star are calculated using these expected phases. Table~\ref{p_vel} shows the largest RV (v$_{\rm max}$) and how much that RV changed ($\Delta$v$_{\rm max}$) for each planet during observations. $\Delta$v$_{\rm max}$ is important as an indicator for the shape of the planetary signal. For example, a planet with a large $\Delta$v$_{\rm max}$ would have broader lines and a lower peak flux, making it harder to detect. Of particular interest are GJ 876b and GJ 876c, which are Jovian-mass planets close to their host star, making them strong candidates for H$_2$ fluorescence signatures. Their RVs during observation were not separable from the host star, which means any H$_2$ emission from the planets would have been included in the stellar flux measurement. This should make the system an outlier when comparing flux ratios. Table~\ref{fluxes} shows that GJ 876 has average emission when compared to other stars in the MUSCLES sample and so we conclude that we do not see any additional signal from the planets. 

\begin{figure}
\centering
\includegraphics[width=170mm,keepaspectratio]{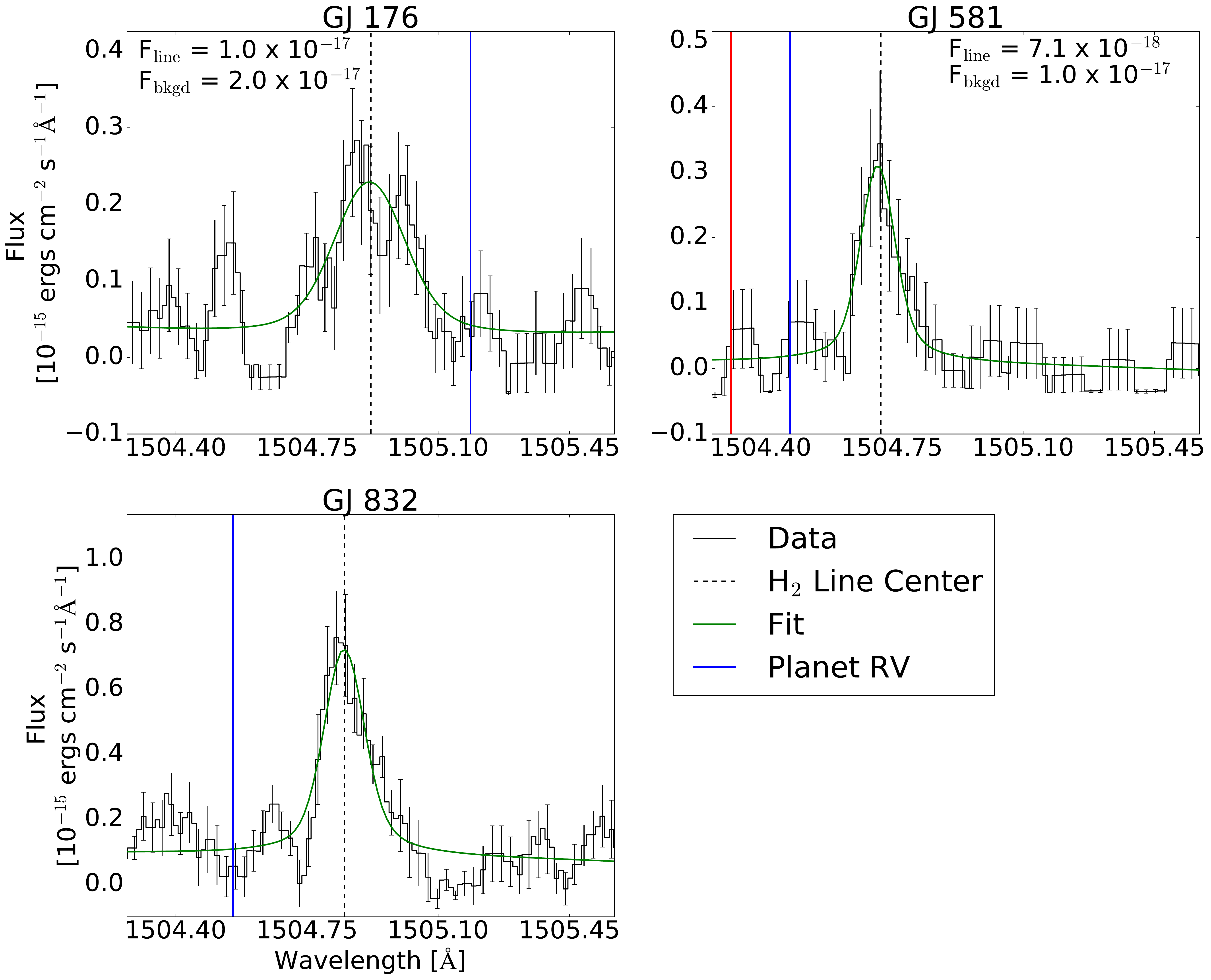}
\caption{Spectra of the brightest fluorescence line for three MUSCLES stars hosting planets with resolvable velocity centroids. The COS observations are shown in black, with the fits from this work overplotted in green. The expected locations of the planetary emission, based off of their RVs in the stellar rest frame, are shown as blue (and red for GJ 581) vertical lines. The largest potential planetary feature for each star was fit and its measured flux (F$_{\rm line}$) is shown} along with the upper limit of the background flux (F$_{\rm bkgd}$). The background flux upper limit was calculated using the procedure described in \S\ref{flares}, but over a region of a few Angstroms instead of $\pm$20 km s$^{-1}$. F$_{\rm line}$ and F$_{\rm bkgd}$ have units of ergs cm$^{-2}$ s$^{-1}$.
\label{planet_em}
\end{figure}

GJ 176, GJ 581 and GJ 832 all have confirmed planets with velocities that were distinguishable from their host stars during the observations. The planets around GJ 628 may have also been distinguishable, but are not included in the following analysis due to a larger uncertainty in their orbit parameters. Figure~\ref{planet_em} shows the (1-7)P(5) transition line for GJ 176, GJ 581, and GJ 832 with vertical lines marking the expected location of the planetary signal. GJ 176 and GJ 581 both have small features that are consistent with the expected line centroids of the planets. However, the flux of these features (F$_{\rm line}$) is less than the upper limit of the background flux (F$_{\rm bkgd}$), showing that the features are consistent with background levels. In all of the cases tested above, planetary H$_2$ fluorescence emission is not seen, ruling out their role as contributors to the observed signal.
\begin{table}
\caption{Orbital parameters for MUSCLES stars with confirmed planets
\label{p_vel}}
\centering
\begin{tabular}{ c c c c c c }
\hline
Star & {Avg. H$_2$ RV \,\tablenotemark{a}} & { $\Delta$RV \,\tablenotemark{b}}& v$_{\rm max}$\,\tablenotemark{c} & $\Delta$v$_{\rm max}$\,\tablenotemark{c} & $i_{\rm max}$\,\tablenotemark{d} \\
& \lbrack km s$^{-1}]$ & \lbrack km s$^{-1}]$ & \lbrack km s$^{-1}]$ & \lbrack km s$^{-1}]$ & \lbrack deg\rbrack \\ \hline
GJ 176 & 29.15 & -7.28 $\pm$ 4.01 & 53.0 & 9.0 & 8.4 \\
GJ 436 & 10.6 & 1.94 $\pm$ 5.7 & -112.3 & 59.0 & 6.1 \\
GJ 581 & -7.2 & -4.99 $\pm$ 4.58 & -48.1, 8.0, & 8.0, 2.0, & 14.1 \\
& & & 24.8, -79.5 & 0.0, 5.0 & \\ 
GJ 628 & -13.9 & 0.64 $\pm$ 3.37 & 73.4, 29.0 & 4.0, 1.0 & 21.8\\
& & & 31.5 & 0.0\\
GJ 667C & 3.3 & 2.49 $\pm$ 6.22 & 41.2, 14.0 & 5.0, 2.0 & 10.1 \\
GJ 832 & 15.1 & -3.12 $\pm$ 4.14 & -0.3, -59.3 & 0.0, 0.0 & 90.0 \\
GJ 876 & 2.5 & -5.09 $\pm$ 4.78 & -11.4, 22.6, & 0.0, 0.0, & 22.4 \\
& & & -82.7, 30.0 & 32.0, 0.0 & \\ \hline
\multicolumn{6}{p{0.6\textwidth}}{$^{\rm a}$ Average RV measured using the two H$_2$ progressions (columns 5 and 6 of Table~\ref{RV})} \\
\multicolumn{6}{p{0.6\textwidth}}{$^{\rm b}$ Largest difference in RV between the velocity measured using the \ion{C}{4} line (column 3 of Table~\ref{RV}) and the two averaged velocities measured using the lines from each H$_2$ progression (columns 5 and 6 of Table~\ref{RV})}  \\
\multicolumn{6}{p{0.6\textwidth}}{$^{\rm c}$ Maximum value during MUSCLES G160M observations, listed alphabetically by planet name} \\
\multicolumn{6}{p{0.6\textwidth}}{$^{\rm d}$ Maximum possible inclination if all H$_2$ fluorescence arises in a disk at the semi-major axis of the planet that is furthest from its host star.}
\end{tabular}
\end{table}

\subsection{Planet-fed Disk Origin} \label{do}
For emission originating in a circumstellar disk, H$_2$ fluorescence would be present in the MUSCLES stars with confirmed planets but not in those without. The emission lines would not have shifted line centers but might have superthermal broadening, depending on the radius and inclination of the disk. If the disk is maintained by an active outflow from a planet, the radius should be linked to the semi-major axis of the planet.

As previously discussed, H$_2$ fluorescence is observed in systems with and without confirmed planets at consistent flux ratios. We find that the H$_2$ RVs are consistent with the stellar RV, but this result is also predicted by the stellar origin hypothesis. To quantify the broadening, we use the larger of the two measured FWHMs (Table~\ref{RV}) of each planet-hosting star to calculate the disk inclination needed to recreate the observed line widths. Under the assumption that the disk follows a Keplarian orbit, the maximum inclination of the disk at the radius of the furthest planet (i$_{\rm max}$) can be calculated using the following equation:

\begin{equation}
\mbox{sin}\;i_{\rm max} = \frac{v_{F}}{2} \sqrt{\frac{R}{GM_{*}}}
\end{equation}
Where $v_{F}$ is the FWHM in km s$^{-1}$, $R$ is the orbital radius of the furthest planet, and $M_{*}$ is the mass of the host star. The resulting distribution of inclinations is compared to the theoretically predicted distribution to find the likelihood that our sample is drawn from a physically realistic population.

The radius of the disk in each system is placed at the semi-major axis of the outer-most planet. This means the calculated $i_{\rm max}$ will be an upper limit. The disk could be fed by a planet on a closer orbit, but this would lead to a higher orbital velocity and therefore a lower inclination angle. Additionally, the measured FWHMs are broadened due to a combination of instrumental effects (e.g., the inherent width of the COS LSF) and the processing of the spectra for analysis (e.g., the coaddition of multiple observations). These narrower FWHMs would require slower apparent velocities at the same orbital radii, meaning that they would need lower inclination angles.

Besides setting the inner edge of the disk, we do not place any constraints on its extent since we are seeing narrow H$_2$ fluorescence lines and it is the inner edge of the disk that determines the maximal broadening. For a 2,500 K population of H$_2$, the thermal broadening is $\sim$4.5 km s$^{-1}$ (where v = $\sqrt{2kT/m}$). This is both smaller than the Keplarian signal we are looking for and smaller than the COS resolution. These comparisons hold out to the dissociation temperature of H$_2$ (T $\sim$ 4,500 K). For these reasons, we do not account for the spatial extent and thermal broadening of the disk in our FWHM and inclination calculations.

The inclination upper limits are listed in Table~\ref{p_vel}. We see that $i_{\rm max}$ for all stars tested, except GJ 832, have values that are less than 23$^{\circ}$. Inclinations follow a probability distribution proportional to sin($i$)~\citep{Lovis10}. We perform a K-S test to compare our results to this expected distribution and find that the two are inconsistent with a confidence $>$99\%. More systems with higher inclinations would be needed to improve the agreement between the two distributions, indicating that our use of upper limits did not affect the results. While this statistic should be taken lightly due to the small number of samples in our survey, the results are supported by the fact that the estimated $i_{\rm max}$ for GJ 436 of 6.1$^{\circ}$ disagrees with the published value of 86.7 $\pm$ 0.03$^{\circ}$~\citep{Knutson11}. Combined with the ubiquity of the emission and consistent flux ratios, these results show that the fluorescence is not originating in a disk maintained by a planetary outflow.

The above analysis does not account for the fact that each star could be hosting an undiscovered planet at an orbit consistent with the measured FWHM. This scenario is ruled out by the direct observations of GJ 832, which show that there is no extra-stellar H$_2$ fluorescence within 10 AU of the star. While a disk could exist outside of this radius, it would likely be too cool to maintain the warm population of H$_2$ needed for the Ly$\alpha$-driven fluorescence to occur. Additionally, the sensitivity of COS begins to fall off outside of $\pm$0.5$^{\prime \prime}$ from the center of the aperture~\citep{Debes16}. Any H$_2$ features further than $\sim$5 AU from the host star would not be seen within the MUSCLES spectroscopic observations.

\subsection{Stellar Origin}
For emission originating near the stellar surface, H$_2$ fluorescence would be seen in systems with confirmed planets and in systems without. The H$_2$ line centers would be consistent with the radial velocity of the star and there would be no superthermal line broadening. The direct imaging results would return no spatially-resolved H$_2$ signal outside of the star itself.

As shown in the discussions on planet (\S\ref{po}) and circumstellar disk origins (\S\ref{po}), we see H$_2$ fluorescence in both populations of MUSCLES stars at comparable flux levels. The measured RVs are consistent with the values obtained from the literature and we do not see significant line broadening. The direct images do not show any sources of H$_2$ emission other than the star.

We also find that 3/4 (one planet host, two non-hosts) of the flare stars have Ly$\alpha$-driven fluorescence. In the situations where no fluorescence was observed, Table~\ref{fluxes} shows that 3-$\sigma$ upper limits on the flux ratios are comparable in size to the measured flux ratios of the other flare stars, indicating that the signal could be present but obscured by the instrumental noise. We also see that the relative line strengths of these stars tend to be smaller than the MUSCLES average. This result can be explained by the increased activity of the flare stars, which causes larger carbon ion luminosities. Taken together, the above evidence shows that a stellar origin for the emission is the most likely scenario.

\section{One-dimensional Spectral Synthesis Modeling} \label{model}
\subsection{H$_2$ Model Description} \label{mod_desc}
We create a simple spectral synthesis model to confirm the stellar hypothesis and gain further insight into the environment in which the H$_2$ resides. We follow the one-dimensional fluorescence modeling procedure of~\cite{McJunkin16} to refine the spatial location of the H$_2$ by recreating the observed H$_2$ fluxes in the MUSCLES stars. We assume that the Ly$\alpha$ emission originates in the chromosphere, where it propagates through a region of neutral hydrogen and deuterium (D/H = 1.5 $\times$ 10$^{-5}$) at a fixed temperature of 10,000 K to a thermalized population of warm H$_2$ near the stellar surface.

We have no knowledge of the geometry of the H$_2$, so we assume a constant column density slab. We only look at the vertical dimension in the model, so the 3-dimensional shape of the H$_2$ and any associated effects, such as limb brightening, are ignored. We restrict our modeling to lines with $\lambda$~$\textgreater$~1350 \AA~so we can ignore any self-absorption of the fluorescent line photons as they leave the emitting region.

For planet-hosting stars, the intrinsic Ly$\alpha$ profiles used in the model are reconstructed from STIS data that was collected as part of the MUSCLES survey. For a detailed description of the reconstruction process, see~\cite{Youngblood16}. For the stars without known planetary systems, no STIS observations were performed and so it is not possible to reconstruct their Ly$\alpha$ profiles. Instead, a proxy is chosen from the available profiles by matching spectral type. The Ly$\alpha$ profile of the best match is then scaled for distance and used as the intrinsic profile for the planetless stars.

The neutral hydrogen and deuterium in the stellar atmosphere attenuate the Ly$\alpha$, limiting the amount that can reach the H$_2$, so the column density of neutral hydrogen ($N_{\rm H\,I}$) is treated as a free parameter in the model. The temperature ($T_{ \rm H_2}$) and column density ($N_{ \rm H_2}$) of the H$_2$ determine how many molecules are available for excitation and so they are also treated as free parameters. All three variables are used to perform a $\chi^2$ minimization between the model and the observed spectrum. A description of the radiative transfer calculations performed within the code can be found in Appendix~\ref{rad_ap}.

A grid search is performed over broad intervals of the three free parameters ($N_{\rm H_2}$= 10$^{14}$-10$^{18}$ cm$^{-2}$, $N_{\rm H\,I}$ = 10$^{12.4}$-10$^{15.8}$ cm$^{-2}$ and $T_{\rm H_2}$ = 1000-5000 K) with finer runs being performed within this range to increase the resolution in an iterative fashion. Because the M dwarf systems display only two H$_2$ progressions, there is a significant degeneracy between the total H$_2$ column density and the ro-vibrational temperature. A lower bound of $T_{ \rm H_2}$ $\gtrsim$ 1500 K can be established by the requirement of warm $v^{\prime \prime}$ = 2 populations. A weakly constrained upper bound in the range $T_{\rm H_2}$ $\lesssim$ 4500 K is set by the H$_2$ collisional dissociation temperature (\citealt{Lepp83}). 

We follow the methods of~\cite{McJunkin16} and calculate the error by holding two parameters fixed while adjusting the remaining parameter until the $\chi^2$ value increased by the 1-$\sigma$ confidence interval. Given the degeneracies described above, these error bars are underestimates of the true uncertainties. However, all best fit temperatures are found within the physically realistic range described above.

\subsection{Model Results} \label{modres}
\begin{figure}
\centering
\includegraphics[width=170mm,keepaspectratio]{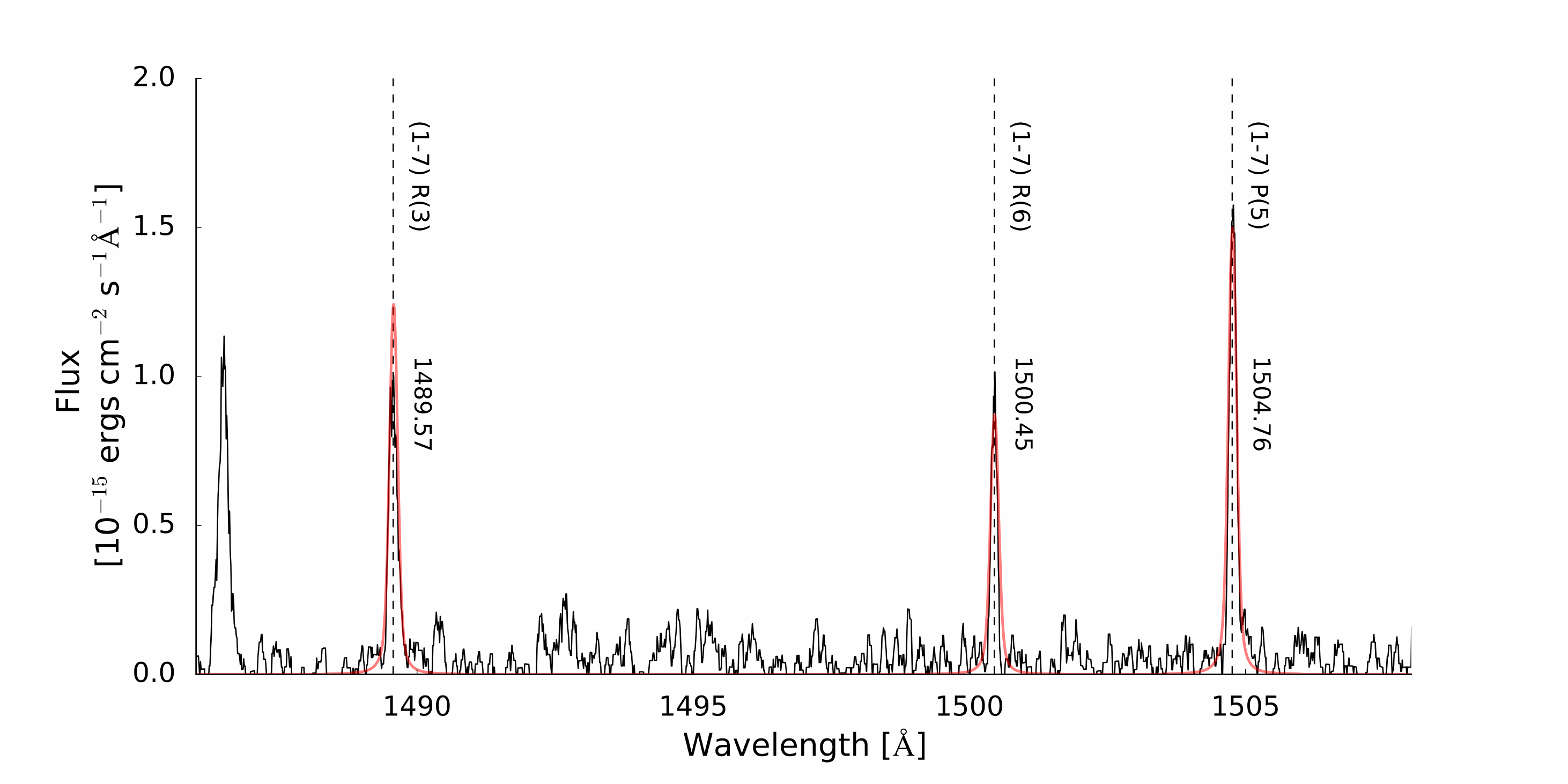}
\caption{A comparison of the MUSCLES spectrum of GJ 876 (black line) to its simulated fluxes (red line). The transition and central wavelength of each H$_2$ fluorescent line have been provided for reference.}
\label{compplot}
\end{figure}

Table~\ref{mres} shows the best fit values and uncertainties for the three free parameters, the effective stellar temperatures ($T_{\rm eff}$), and $\chi_{\nu}^2$ for each model. Figure~\ref{compplot} shows the simulated spectrum compared to the observed data for an example M dwarf. We find that an $N_{\rm H\,I}$ $\gtrsim$ 10$^{14.2}$ cm$^{-2}$ is a breaking point, above which the Ly$\alpha$ flux is so heavily attenuated that not enough remains to excite the observed flux in the [1,7] progression. GJ 628 and GJ 887 have higher than expected $\chi_{\nu}^2$ values of 7.62 and 5.81, respectively. A possible explanation is that the Ly$\alpha$ proxy profiles (Section~\ref{mod_desc}) of both of these stars do not adequately describe their true exciting radiation fields.

\begin{table}
\caption{Model best fit parameters
\label{mres}}
\centering
\begin{tabular}{ c c c c c c c c }
\hline
Star & log$_{10}$($N_{\rm H_2}$) & log$_{10}$($N_{\rm H\,I}$) & $T_{\rm H_2}$ & $T_{\rm eff}$ & Ref. & $\chi_{\nu}^2$ & D.O.F \\
& log$_{10}$([cm$^{-2}$]) & log$_{10}$([cm$^{-2}$]) & \lbrack K\rbrack & \lbrack K\rbrack & & & \\ \hline
GJ 176 & 16.43$_{-0.05}^{+0.07}$ & 13.81$_{-0.18}^{+0.20}$ & 1920$_{-40}^{+48}$ & 3416 $\pm$ 100 & 1 & 0.68 & 5\\
GJ 436 & 16.12$_{-0.06}^{+0.07}$ & 14.30$_{-0.09}^{+0.07}$ & 2440$_{-80}^{+86}$ & 3281 $\pm$ 110 & 1 & 1.13 & 2\\
GJ 581 & 15.92$_{-0.08}^{+0.08}$ & 14.20$_{-0.08}^{+0.07}$ & 3750$_{-274}^{+252}$ & 3295 $\pm$ 140 & 1 & 1.19 & 4\\
GJ 628 & 15.90$_{-0.02}^{+0.02}$ & 14.22$_{-0.02}^{+0.02}$ & 2810$_{-38}^{+36}$ & 3570 & 2, 3 & 7.62 & 9 \\
GJ 667C & 15.70$_{-0.13}^{+0.18}$ & 14.19$_{-0.15}^{+0.11}$ & 2620$_{-186}^{+223}$ & 3327 $\pm$ 120 & 1 & 3.10 & 1 \\
GJ 832 & 15.52$_{-0.05}^{+0.05}$ & 14.00$_{-0.07}^{+0.07}$ & 2755$_{-80}^{+81}$ & 3816 $\pm$ 250 & 1 & 2.05 & 6 \\
GJ 876 & 17.14$_{-0.02}^{+0.02}$ & 14.01$_{-0.03}^{+0.03}$ & 1925$_{-13}^{+14}$ & 3062$_{-130}^{+120}$ & 1 & 1.21 & 10 \\
GJ 887 & 15.99$_{-0.03}^{+0.03}$ & 14.29$_{-0.04}^{+0.04}$ & 2750$_{-42}^{+45}$ & 3676 $\pm$ 35 & 3, 4 & 5.04 & 9\\
GJ 1061 & 15.47$_{-0.09}^{+0.11}$ & 14.02$_{-0.12}^{+0.10}$ & 3490$_{-282}^{+279}$ & 2879 & 5 & 0.67 & 2 \\
HD 173739 & 15.64$_{-0.03}^{+0.03}$ & 14.21$_{-0.04}^{+0.03}$ & 2820$_{-47}^{+45}$ & 3407 $\pm$ 15 & 3, 4 & 0.92 & 7 \\ \hline \hline
Average & 16.34 $\pm$ 0.44 & 14.15 $\pm$ 0.42 & 2728 $\pm$ 550 & --- & --- & --- & --- \\ \hline
\multicolumn{7}{p{0.65\textwidth}}{{\bf Note:} Stars with two references list the original reference for the value and the catalog from which it was obtained.} \\
\multicolumn{7}{p{0.65\textwidth}}{References for T$_{\rm eff}$: 1)~\cite{Loyd16}; 2)~\cite{Zboril98}; 3)~\cite{Soubiran16}; 4)~\cite{Boyajian12}; 5)~\cite{Gautier07}} \\
\\
\end{tabular}
\end{table}

The location of the emission is further constrained using the results of our model. For eight out of the ten MUSCLES stars with observed H$_2$ emission, the best fit $T_{\rm H_2}$ is lower than $T_{\rm eff}$. This cooler emission temperature has two possible explanations. The first is that the emission is originating in starspots. This is supported by the average $T_{\rm H_2}$/$T_{\rm eff}$ of 0.83, which roughly agrees with the theoretical predictions of~\cite{Jackson13} who found a spot-to-photosphere temperature ratio of 0.70 $\pm$ 0.05. Another possible explanation is that the molecules are dispersed within a cooler layer of the star. The semi-empirical stellar atmosphere model of GJ 832 from~\cite{Fontenla16} shows that the lower chromosphere can reach a temperature minimum of $\sim$2650 K, which agrees well with our modeled T$_{\rm H_2}$ of 2755 $\pm$ 80 K. Given the range of spectral types in the MUSCLES sample, we consider a range of lower chromosphere temperatures between 2500-3000 K to be reasonable.

Our modeled H$_2$ column densities are smaller than values obtained previously for planetary nebulae and protoplanetary disks (\citealt{Herczeg04, Lupu06, Schindhelm12}). A possible explanation for lower values in the M dwarf atmospheres is that the observed fluorescence is occurring in a thin layer below which physical conditions are no longer favorable for H$_2$ fluorescence. As hydrogen density increases with higher pressures, the available Ly$\alpha$ flux is rapidly extinguished, removing the exciting photons. The other possibility is that, with the sharp increase in temperature in the upper chromosphere, the H$_2$ begins to collisionally dissociate thereby limiting the fluorescence process.

\section{CONCLUSIONS} \label{conclude}
In this work, we studied Ly$\alpha$-driven H$_2$ fluorescence, which was previously observed in M dwarf systems but without sufficient detail to determine its spatial origin. We searched the {\it HST} spectra of 11 M dwarfs, which were observed as part of the MUSCLES Treasury Survey, and four previously observed, highly-active M dwarfs, for signs of H$_2$ fluorescence. We compare the velocity centroids, line widths, and relative strengths of the H$_2$ emission features in M dwarfs with and without known planets to determine the origin of the fluorescence. The results are further supported by the direct imaging of the GJ 832 system.\\
\\
Our combined analyses show that the H$_2$ fluorescence has a stellar origin:
\begin{enumerate}
\item{Fluorescence is seen in stars with and without known planets (Table~\ref{h2num}).}
\item{H$_2$ radial velocities are consistent with the stellar radial velocity and no superthermal broadening (indicative of rotation in a circumstellar disk) is observed (Table~\ref{RV}).}
\item{We find that the H$_2$-to-stellar-ion flux ratios, which are used to compare the relative H$_2$ fluxes amongst the stars, are consistent between planet hosts and non-hosts. This indicates that the planets are not contributing to the observed signal (Table~\ref{fluxes}).}
\item{No exoplanetary or circumstellar disk emission is observed in the direct FUV imaging of GJ 832 (Figure~\ref{cosovly}).}
\item{Estimated inclinations needed to reproduce observed FWHMs are inconsistent with the predicted distribution indicating that our sample does not reflect a physically realistic population of circumstellar disks (Table~\ref{p_vel}).}
\end{enumerate}
A new analysis shows that H$_2$ emission is detected in three out of the four previously observed active M dwarfs. This weak fluorescent signal was missed in previous analyses as a result of the larger instrumental noise floor of {\it HST}-STIS.

Our radiative transfer model provides further constraints on the location of the fluorescence. Based on the modeled temperatures, it is possible that the H$_2$ resides either in starspots or in a cooler region of the lower chromosphere. Our modeled column densities are lower than what is observed in other systems, such as protoplanetary disks, indicating that the fluorescence may only be occurring in a thin region on the stars. We find that, while H$_2$ fluorescence could be a powerful tool for characterizing exoplanetary atmospheres in the future, the fact that the emission is associated with the star could adversely affect future attempts at observing it in exoplanet atmospheres. In some cases, phase-resolved observations made with high-resolution FUV spectroscopy should be sufficient to extract putative planetary signals from the emission arising near the stellar surface, although H$_2$ fluorescence may not be visible in many cases, as was shown for GJ 176, GJ 581 and GJ 832 (Figure~\ref{planet_em}). However, H$_2$ emission resulting from electron impacts in giant planet aurorae (\citealt{France10,Gustin12}) would not be similarly compromised since it would not have an equivalent stellar source.

\section{ACKNOWLEDGEMENTS}
The data presented here were obtained as part of the HST Guest Observing programs \#13650 and \#14100. RV calculations for GJ 1061 are based on observations collected at the European Organisation for Astronomical Research in the Southern Hemisphere under ESO programme 089.C-0904(A).

\appendix 

\section{Stellar Observations of GJ 832 with the ACS/SBC} \label{obs_ap}

The flux seen in the ACS/SBC FUV filter observations is dominated by emission lines from the stellar chromosphere and transition region (e.g., F13). We use our existing spectroscopic observations of GJ 832 to quantify the contribution of a given line to each photometric flux value and explore time variability in the upper atmosphere of GJ 832. Fluxes measured using the ACS/SBC observations compare well with spectroscopically derived fluxes from $HST$-STIS (Ly$\alpha$, for comparison with the F122M) and $HST$-COS (1350-1700~\AA, for comparison with F140LP) MUSCLES observations. Given the red leak in the longer wavelength long-pass filters in the ACS/SBC, we do not attempt to compare the F165LP fluxes with our spectroscopic data\footnote{We note that the red leak for cool stars is overestimated in the STScI ACS Handbook because the Pickles stellar models do not include realistic estimates of the UV-bright chromospheres of G, K, and M dwarfs.}. The F122M (Ly$\alpha$) flux in a 1\arcsec\ photometric aperture is $F_{\rm phot}$(122M) = 1.9~$\times$~10$^{-13}$ erg cm$^{-2}$ s$^{-1}$, compared with the observed Ly$\alpha$ emission line profile (integrated over 1214.5-1217.5~\AA) of $F_{\rm COS}$(122M) = 2.5~$\times$~10$^{-13}$ erg cm$^{-2}$ s$^{-1}$ (a 24\% difference). The F140LP (H$_{2}$ and other chromospheric and transition region emission lines) flux in a 1\arcsec\ photometric aperture is $F_{\rm phot}$(140LP) = 3.4~$\times$~10$^{-14}$ erg cm$^{-2}$ s$^{-1}$, compared with the observed $HST$-COS spectrum (integrated over 1380-1680~\AA) of $F_{\rm COS}$(140LP) = 3.8~$\times$~10$^{-14}$ erg cm$^{-2}$ s$^{-1}$ (an 11\% difference). M dwarfs exhibit variations in the chromospheric Ly$\alpha$ emission line flux of 10-30\% on timescales of a few years (see, e.g.,~\citealt{Youngblood16}), consistent with the photometric and spectroscopic comparison of GJ 832 presented here. The stellar flux is consistent with no change over the approximately six weeks between visits, suggesting that no significant flares occurred during either of the ACS/SBC observations. 

Comparing the fluxes of the F122M and the F140LP bands, we observe that the strength of the Ly$\alpha$ emission from a typical M dwarf like GJ 832 is a factor of 5-6 greater than the longer-wavelength FUV emission, even though the Ly$\alpha$ emission profile has not been corrected for attenuation by interstellar neutral hydrogen. We estimate the relative contributions of each of the major spectral features to the ACS/SBC flux by comparing the flux in the F140LP emission line with the archival $HST$-COS spectra: \ion{Si}{4} (9.5\%), \ion{C}{4} (34.9\%), \ion{He}{2}
(21.1\%), and the 1430-1520 \AA\ H$_{2}$ region (14.6\%). The remaining flux is contributed from the many weak neutral lines (mainly \ion{S}{1} and \ion{N}{1}) and a FUV continuum emission (see~\citealt{Loyd16}) in the F140LP bandpass. 

\section{H$_2$ Radiative Transfer Calculations} \label{rad_ap}
The H$_2$ fluorescence code calculates the observed flux in each of the two transitions using the formulae outlined below (see also, \citealt{McJunkin16}). First, the oscillator strengths for absorption ($f_{\rm lu}$) are calculated using Einstein A coefficients ($A_{\rm ul}$) from~\cite{Abgrall93} and the degeneracies of the upper and lower states ($g_{\rm u}$ and $g_{\rm l}$, respectively)
\begin{equation}
f_{\rm lu} = \frac{m_e c}{8 \pi^2 e^2} \frac{g_{\rm u}}{g_{\rm l}} \lambda^2_{\rm lu} A_{\rm ul}
\end{equation}
The reconstructed Ly$\alpha$ profiles (F$_{\rm Ly\alpha}$) from~\cite{Youngblood16} are attenuated by a population of H atoms (parameterized by $N_{\rm H I}$) between the source of the Ly$\alpha$ photons and the H$_2$ slab. The remaining flux pumps a thermally-populated ensemble of H$_2$ along one of the two progressions of interest. The absorbing cross-sections of the molecules are calculated using
\begin{equation}
\sigma_{\rm lu} = \frac{\sqrt{\pi}e^2}{m_e c b} f_{\rm lu} \lambda_{\rm lu} H(a,y)~~~~~\mbox{[cm$^{2}$]}
\end{equation}
where $H(a,y)$ is the Voigt function with
\begin{equation}
a = \frac{\Gamma}{4\pi \Delta \nu}
\end{equation}
\begin{equation}
y = \frac{|\nu - \nu_0|}{\Delta \nu}
\end{equation}
\begin{equation}
\Delta \nu = \frac{b}{c} \nu
\end{equation}
and $\Gamma$ values obtained from~\cite{Abgrall93}. These cross-sections are used to calculate an optical depth
\begin{equation}
\tau = N(v,J)\sigma_{\rm lu}
\end{equation}
where $N(v,J)$ is
\begin{equation}
N(v,J) = N_{\rm H_2} \frac{(2J+1)(2s+1)e^{E(v,J)/k_b T_{\rm H_2}}}{\Sigma_{v,J} (2J+1)(2s+1)e^{E(v,J)/k_b T_{\rm H_2}}}~~~~~\mbox{[cm$^{-2}$]}
\label{coldens}
\end{equation}
where s is the nuclear spin (0 for even J-values, 1 for odd J-values), $E(v,J)$ is the energy in each $(v,J)$ state, and $N_{\rm H_2}$ and $T_{\rm H_2}$ are the column density and temperature parameters that are passed into the model. The flux absorbed by each transition is then given by integrating over the full Ly$\alpha$ line
\begin{equation}
F_{\rm abs} = \int (1 - e^{-\tau}) F_{\rm Ly\alpha}~~~~~\mbox{[erg cm$^{-2}$ s$^{-1}$]}
\end{equation}
The flux in each of the two upper states is then redistributed amongst the fluorescent lines of the given progression. The flux in a given H$_2$ fluorescent line is calculated by multiplying the total progression flux by the branching ratio of the fluorescent transition. The branching ratio is given by
\begin{equation}
r_{\rm branch} = \frac{A_{\rm ul}}{\Sigma A_{\rm ul}}
\end{equation}
The resulting fluxes are used in the $\chi^2$ minimization, along with the fluxes calculated from the MUSCLES spectra (\S\ref{model}).

\bibliography{H2Flo_bib}

\end{document}